\documentclass[reveiw, 3p, times, 11pt]{elsarticle}

\usepackage{amssymb}

\usepackage{mathtools}
\usepackage{xcolor}
\usepackage{booktabs}
\usepackage{pifont}

\usepackage{tikz}
\usepackage{enumitem}
\usetikzlibrary{arrows.meta}
\usepackage{caption} 
\usetikzlibrary{shapes, snakes, calc}
\pgfdeclarelayer{background}
\pgfdeclarelayer{foreground}
\pgfsetlayers{background, main, foreground}
\usepackage{lipsum}
\usepackage{amsmath}
\usepackage{contour}
\definecolor{mygray}{HTML}{dcddde}
\usepackage[utf8]{inputenc} 
\usepackage[T1]{fontenc} 
\usepackage{helvet}
\usepackage{url}
\usepackage{layout}

\usepackage{array}
\usepackage{colortbl}
\usepackage{xcolor}
\usepackage{hhline}
\usepackage{float}
\usepackage{multirow}
\usepackage{booktabs}
\usepackage{ragged2e}
\newcolumntype{C}[1]{>{\centering\arraybackslash} p{#1}}
\newcolumntype{T}[1]{>{\RaggedRight\arraybackslash}m{#1}}
\newcolumntype{F}[1]{@{\hspace{5pt}}T{#1}}
\newcolumntype{L}[1]{T{#1}@{\hspace{3pt}}}
\usepackage{hyperref} 
\hypersetup{colorlinks=true, breaklinks=true, bookmarksopen=false,}
\definecolor{darkgreen}{rgb}{0.0, 0.5, 0.0}    
\definecolor{darkblue}{rgb}{0.0, 0.0, 0.5}     
\definecolor{slategray}{rgb}{0.44, 0.5, 0.56}  

\makeatletter
    \AtBeginDocument{\def\@citecolor{darkgreen}}
    \AtBeginDocument{\def\@linkcolor{darkblue}}
    \AtBeginDocument{\def\@urlcolor{blue}}
\makeatother

\usepackage[acronym,nonumberlist]{glossaries}
\makeglossaries
\newacronym{NECP}{NECP}{National Energy and Climate Plan}
\newacronym{GHG}{GHG}{Greenhouse Gases}
\newacronym{ETS}{ETS}{Emissions Trading System}
\newacronym{STEP}{STEP}{Social, Technological, Economic, and Political}
\newacronym{STEEP}{STEEP}{Social, Technological, Economic, Environmental, and Political}
\newacronym{STEP+G}{STEP+G}{STEP + Geopolitics}
\newacronym{Q2Q}{Q2Q}{Qualitative to Quantitative}
\newacronym{IEA}{IEA}{International Energy Agency}
\newacronym{RD}{{R\&D}}{Research and Development}
\newacronym{EC}{EC}{European Commission}
\newacronym{EU}{EU}{European Union}
\newacronym{MS}{MS}{Member-State}
\newacronym{NDC}{NDC}{Nationally Determined Contributions}
\newacronym{LTS}{LTS}{Long-Term Strategies}
\newacronym{EnVis}{EU EnVis-2060}{European Energy Vision 2060}
\newacronym{WEM}{WEM}{With Existing Measures}
\newacronym{WAM}{WAM}{With Additional Measures}
\newacronym{GECO}{GECO}{Global Energy and Climate Outlook}
\newacronym{IPCC}{IPCC}{Intergovernmental Panel on Climate Change}
\newacronym{IRENA}{IRENA}{International Renewable Energy Agency}
\newacronym{WEC}{WEC}{World Energy Council}
\newacronym{CCU}{CCU}{Carbon Capture and Utilization}
\newacronym{RER}{RER}{Renewable Energy Resources}
\newacronym{ESS}{ESS}{Energy Storage Solutions}
\newacronym{CCS}{CCS}{Carbon Capture \& Storage}
\newacronym{DACS}{DACS}{Direct Air Capture and Storage}
\newacronym{TYNDP}{TYNDP}{Ten-Year Network Development Plan}
\newacronym{SSP}{SSP}{Shared Socioeconomic Pathway}

\makeatletter
\newcommand\notsotiny{\@setfontsize\notsotiny\@vipt\@viipt}
\makeatother

\journal{arXiv. This manuscript is a preprint and has not yet been peer-reviewed.}

\begin{document}

\begin{frontmatter}

\title{\LARGE European Energy Vision 2060: \\Charting Diverse Pathways for Europe’s Energy Transition}

\author[1]{Mostafa~Barani\corref{mycorrespondingauthor}}
\cortext[mycorrespondingauthor]{Contact Information:}
\ead{mostafa.barani@ntnu.no}
\author[2,1]{Konstantin~Löffler}
\author[1]{Pedro~Crespo~del~Granado}
\author[2]{Nikita~Moskalenko}
\author[3]{Evangelos~Panos}
\author[4,8]{Franziska~M.~Hoffart}
\author[2]{Christian~von~Hirschhausen}
\author[5]{Maria~Kannavou}
\author[6,1]{Hans~Auer}
\author[2]{Karlo~Hainsch}
\author[1]{Tatiana~González~Grandón}
\author[7]{Siri~Mathisen}
\author[1]{Asgeir~Tomasgard}

\affiliation[1]{organization={Department of Industrial Economics and Technology Management, Norwegian University of Science and Technology~(NTNU)},
            city={Trondheim},
            country={Norway}}

\affiliation[2]{organization={Workgroup for Infrastructure Policy, Technische Universität Berlin~(TU~Berlin)},
            city={Berlin},
            country={Germany}}

\affiliation[3]{organization={Energy Economics Group, Laboratory for Energy Systems Analysis, PSI Centers for Nuclear Engineering \& Sciences and Energy \& Environmental Sciences},
            city={VIlligen PSI},
            country={Switzerland}}

\affiliation[4]{organization={Kassel Institute for Sustainability, University of Kassel},
            city={Kassel},
            country={Germany}}

\affiliation[8]{organization={Sociological Research Institute Goettingen e.V., University of Goettingenl},
            city={Goettingenl},
            country={Germany}}

\affiliation[5]{organization={E3-Modelling~SA.~(A~Ricardo~Company)},
            city={Athens},
            country={Greece}}

\affiliation[6]{organization={Energy Economics Group (EEG), Technische Universität Wien~(TU~Wien)},
            city={Vienna},
            country={Austria}}

\affiliation[7]{organization={Department of Energy Systems, SINTEF Energy Research},
            city={Trondheim},
            country={Norway}}

\begin{abstract}

Europe is warming at the fastest rate of all continents, experiencing a temperature increase of about 1°C higher than the corresponding global increase. Aiming to be the first climate-neutral continent by 2050 under the European Green Deal, Europe requires an in-depth understanding of the potential energy transition pathways. 
In this paper, we develop four qualitative long-term scenarios covering the European energy landscape, considering key uncertainty pillars—categorized under social, technological, economic, political, and geopolitical aspects. 
First, we place the scenarios in a three-dimensional space defined by Social dynamics, Innovation, and Geopolitical instabilities. These scenarios are brought to life by defining their narratives and focus areas according to their location in this three-dimensional space. The scenarios envision diverse futures and include distinct features. The \textit{EU Trinity} scenario pictures how internal divisions among EU member states, in the context of global geopolitical instability, affect the EU climate targets. The \textit{REPowerEU++} scenario outlines the steps needed for a self-sufficient, independent European energy system by 2050. The \textit{Go RES} scenario examines the feasibility of achieving carbon neutrality earlier than 2050 given favourable uncertain factors. The \textit{NECP Essentials} scenario extends current national energy and climate plans until 2060 to assess their role in realizing climate neutrality. The scenarios are extended by incorporating policies and economic factors and detailed in a Qualitative to Quantitative (Q2Q) matrix, linking narratives to quantification. Finally, two scenarios are quantified to illustrate the quantification process. All the scenarios are in the process of being quantified and will be openly available and reusable.
\end{abstract}


\end{frontmatter}

\clearpage

\printglossaries

\clearpage

\section{Introduction}

The alarm bells of climate change are ringing louder than ever. We are witnessing an alarming acceleration of the ecological crisis and climate emergency, marked by record-breaking high temperatures and extreme weather events attributed to climate change worldwide \cite{jones2022a, dong2024extreme}. The COP28 stocktake in December 2023 indicated that global efforts are insufficient to meet the Paris Agreement's objective of limiting the rise in global temperature to 1.5°C, making it imperative for nations to exhibit significant progress by 2030 to avert further climate degradation \cite{UFCCC}.

The European nations are no exception. In fact, the \gls{EU}, aiming to become the first climate-neutral continent by 2050 as outlined in the European Green Deal \cite{EuropeanGreenDeal}, stands at a critical juncture in its energy policy landscape. This transition is driven by the urgent need to achieve ambitious climate targets while navigating a complicated geopolitical landscape \cite{goldthau2023eu}. Achieving such ambitious goals necessitates deep insights into the future energy landscape in Europe, as well as planning  appropriate policies and strategies.

In response, the European Commission has unveiled comprehensive strategies and policies to address both immediate and long-term energy needs. One of these strategies is the REPowerEU plan, launched to bolster the \gls{EU}'s energy independence and accelerate the shift towards renewable energy sources in light of geopolitical disruptions caused by Russia's invasion of Ukraine \cite{REPowerEU}. This initiative works in tandem with the \textit{Fit for 55} package, which seeks to reduce net \gls{GHG} emissions by at least 55\% by 2030 \cite{Fitfor55}, and the Clean Energy for All Europeans package, which requires each EU member state to deliver 10-year \gls{NECP} plans \cite{CleanEnergy}. The \textit{Fit for 55} goals include a reduction in \gls{ETS} and non-\gls{ETS} sectors with 61\% and 40\%, respectively, compared to 2005 levels \cite{Fitfor55}.


 Navigating the complexities of policy-making and planning, especially over the long term, is significantly challenged by the uncertainties (e.g. geopolitical, technological adoption and breakthroughs) that shape energy transitions. To address these challenges, scenario techniques offer a strategic framework that fosters insight and supports decision-making in the face of deep uncertainty \cite{bishop2007a}. By exploring various pathways for energy transition, scenarios take into account a range of factors, including technological advancements, societal trends, and political dynamics. The European Commission has long utilized scenario analysis (e.g., \cite{2020EURef,JRC136265,CT2040}) to evaluate the potential impacts of different actions, guiding amendments to existing directives, setting climate and energy targets, and introducing new policies.

Due to their complexity, long-term energy scenarios typically include two elements: qualitative and quantitative. The qualitative element consists of narrative descriptions of possible future events, conditions, or developments across various dimensions, such as society \cite{van2005a}. The quantitative element, derived from the qualitative narratives, involves translating these descriptions into numerical data. This process is not standardized and often relies on mathematical models, typically energy system models in this context.

Despite the existence of numerous long-term European energy scenarios (see Section \ref{sec:LiteratureReview}), the world is not static and unchanging. New insights regarding evolving uncertainties continually emerge. For instance, recent years have seen disruptions and tensions that have established a more fractured world within which the EU strives to implement its decarbonization agenda. Concurrently, the pace of disruptions has accelerated. These include rapid shifts in societal norms—such as declining ownership and the rise of circular and sharing economies—increasing awareness of climate change, acceleration of decentralized electrification, convergence of new technologies, and exponential data accumulation. In such a dynamic and rapidly changing environment, the regular development of energy scenarios is essential for exploring and navigating these emerging energy dynamics. 

This paper develops qualitative energy scenarios that cover \gls{EU} until 2060.
It makes several contributions to the existing literature on energy scenario development and long-term energy scenarios:
First, it provides a comprehensive overview of long-term European energy transition scenarios, identifying relevant challenges and opportunities and offering insights into the status of the EU energy transition over time.
Second, it contributes methodologically by more effectively engaging experts' ideas (including those of stakeholders and need owners) in qualitative scenario development; adding a companion, called ``Areas of Focus,'' to each scenario narrative to make it more engaging; combining the 3-D space method in \cite{openentrance1} with Morphologic analysis to include more uncertainty factors in the scenario definitions; and proposing a detailed table called the \gls{Q2Q} matrix to partially systemize the translation of qualitative scenarios into quantitative counterparts.
Third, it introduces four qualitative European energy transition scenarios, situated within a specific three-dimensional uncertainty space where each dimension represents a key uncertainty, with current geopolitical instabilities highlighted as one of its core dimensions. The scenarios are then detailed through comprehensive \textit{narratives} and \textit{focus areas}.
Finally, two of the scenarios are quantified, illustrating the translation of a qualitative scenario into a quantitative counterpart through an optimization-based energy transition modelling tool. 
This set provides a detailed framework for scenario analysis and serves as a tool for researchers and policymakers to explore and plan for future energy systems.


The remainder of this paper is organized as follows: Section \ref{sec:LiteratureReview} presents a comprehensive literature review on scenario development techniques and energy transition scenarios, with a particular emphasis on the EU context. Section \ref{sec:qualitative} details the four qualitative scenarios developed in this study, along with the methodologies employed. Section \ref{sec:quantitave} illustrates the quantification process by converting one of the qualitative scenarios into a quantitative model. Finally, Section \ref{sec:conclusion} provides conclusions and directions for future research.

\section{Literature Review} \label{sec:LiteratureReview}
This section presents a detailed review of the processes in defining scenarios and types of scenarios in general. The review summarizes the most relevant European and global energy transition scenarios along with practices of stakeholder associations and energy companies. The section concludes with a brief discussion of solely qualitative energy transition scenarios. In addition to the description of the individual scenario studies, the main characteristics and features of these studies are compared in Tables \ref{tab:literature1}, \ref{tab:literature2}, and \ref{tab:literature3}.
\subsection{Definition and Types of Scenarios}

Scenario analysis has a long tradition \cite{smil2000perils} and was first used for military strategies \cite{kahn1967year} and in the late 1960s in the energy and business context by Royal Dutch Shell \cite{andersson2020ghost}. Today, they are central tools in climate change and energy transition research, where uncertainty and complexity are high \cite{poganietz2020introduction, pregger2020moving, hoffart2021_1}. In the literature, there is no unified standard scenario approach; instead, there exist divergent methods, typologies and definitions \cite{bradfield2005origins}. Generally, a scenario can be described as “a possible situation in the future, based on a complex network of influence factors” \cite{gausemeier1998scenario}. It displays “different assumptions about how current trends will unfold, how critical uncertainties will play” \cite{kosow2008methods}.

Scenario typologies are distinguished between (i) predictive, (ii) exploratory and (iii) normative scenarios,  asking (i) what will happen, (ii) what might happen, and (iii) how can a normative target be reached respectively \cite{borjeson2006scenario}. The former is also called forecasting and mostly consists of trend analysis, which extrapolates from past and present to the one most probable future using data and simulations. The latter two, also known as foresight qualitative scenarios, are key-factor based scenario techniques using either systematic-formalized, such as consistency or cross impact analysis, or creative narrative approaches, such as normative narrative or morphological analysis \cite{kosow2008methods}.

\begin{figure}
    \centering
    \includegraphics{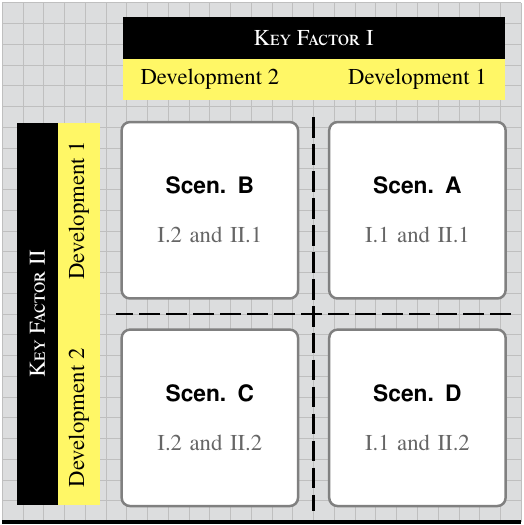}
    \caption{Scenario development technique: Reducing the system to only two key factors, each with two possible developments, resulting in four distinct scenarios.}
    \label{fig:twobytwomethod}
\end{figure}

The scenario development techniques applied thus far have been heavily reliant on the expertise of the team responsible for constructing the scenarios. Historically, the most simplified approaches were either based on a reference scenario, supplemented by variations in key parameters to generate additional scenarios, or focused on reducing the essential factors to only two. The latter method, widely adopted in the context of energy scenarios, considers two possible developments for each of these two factors, resulting in four distinct scenarios, as illustrated in Figure \ref{fig:twobytwomethod}. The narratives are subsequently developed for each scenario.

\subsection{EU Energy Transition Scenarios: European Commission}
The \gls{EC} has been using scenario analysis for many years to support decision-making \cite{EU17}. This approach helps to explore and evaluate the interactions between energy, the economy, and the environment. It is used to assess impacts that support proposals to amend existing directives, set climate and energy targets, or introduce new ones. Typically, it includes a reference scenario and additional scenarios that incorporate changes in various dimensions, such as climate and energy policies, technological progress, and more. The scenarios are constructed around narratives or storylines that describe how different combinations of factors might unfold, representing plausible future trajectories.

\subsubsection{EU Reference Scenario}
The \gls{EU} Reference Scenario has been a cornerstone of the European Commission's analytical toolkit for over two decades, offering insights into energy, transport and climate action. This scenario, a projection (not a forecast) of the \gls{EU} energy system, transport system, and GHG emissions, is a pivotal baseline for comparing policy scenarios and drawing policy recommendations. It enables policymakers to delve into the long-term economic, energy, climate and transport outlook within the existing policy framework, providing a robust analytical basis for assessing new policy proposals. The scenario reflects current market and technology trends and national or EU-wide policies to be implemented with a specific cut-off date. The 2020 EU Reference Scenario \cite{2020EURef} accounts for all policies and targets included in the 2019 \gls{NECP} \gls{WAM} scenarios and other plans and measures put forward as of the end of 2019.

The EU Reference Scenario is updated every 3 to 4 years, with the most recent version published in 2020 and other recent versions in 2016 \cite{2016EURef} and 2014 \cite{2013EURef}. The updates incorporate the latest available statistics on the macro-economy, demographics, and energy trends, as well as projections for future years. They also consider developments such as changes in policies, energy and climate targets (e.g., updated \glspl{NECP} and \gls{MS}-specific policies and targets, new \gls{EU} legislative measures), technology (e.g., the impact of the ``fracking revolution'' on fossil fuel prices), and international geopolitical events (e.g., the COVID-19 impact on the economy and energy trends, the Fukushima nuclear accident) that influence these projections. 

\subsubsection{2040 Climate Target Plan Scenarios}
Building on the foundation of the \gls{EU} Reference Scenario, the European Commission in early 2024 recommended that the \gls{EU}'s net \gls{GHG} emissions be reduced by 90\% by 2040 compared to 1990 levels. This marked the first time a 2040 climate target was set. To establish these targets, the European Commission conducted a comprehensive impact assessment \cite{CT2040} that evaluated various levels of GHG emission reduction for 2040. The European Commission developed different policy scenarios to define the 2040 target, aiming to meet objectives such as ensuring a fair transition to a green economy and climate neutrality by 2050 while maintaining the cost-competitiveness of the European economy. The policy scenarios include three core scenarios (S1, S2, S3) and a variant (LIFE) that demonstrates how behavioral and lifestyle changes in a circular economy can complement overall efforts to reduce GHG emissions.

All these scenarios share the same key assumptions about the macro-economy, demographics, and international energy markets. They also have the same starting and ending points regarding climate and energy goals, specifically the 2030 target of reducing \gls{GHG} emissions by 55\%, other associated energy and climate targets outlined in the \textit{Fit-for-55} policy framework, and the 2050 climate neutrality goal. The scenarios differ in the post-2030 \gls{GHG} reduction rate, reflecting the contributions from various sectors and technologies. The carbon budget for 2040 varies across different scenarios: from S1, which allows for a ``linear'' reduction path between 2030 and 2050, to S2, a more ambitious scenario aiming for an 85\% \gls{GHG} emissions reduction by 2040, and S3, which builds on S2 and, through a fully developed carbon management industry, achieves a 90\% net \gls{GHG} emissions reduction by 2040.

\begin{table*}
    \centering
    \scriptsize
    \caption{Reviewed scenarios at a glance -- Part 1 of 3.}
    \makebox[\textwidth][c]{%
    \resizebox{1\textwidth}{!}{
    \begin{tabular}{@{}F{35pt}T{50pt}T{70pt}T{50pt}T{23pt}T{50pt}T{35pt}L{80pt}@{}}
    \specialrule{2pt}{0pt}{0pt}\\[-4pt]
            \textbf{Name} & \textbf{Conducted By} & \textbf{Scenarios} & \textbf{Temp. Rise} & \textbf{Horizon} & \textbf{Coverage} & \textbf{Dev. Date} & \textbf{Key Considerations}   \\[5pt]
            \cmidrule[1pt]{1-8}
            open ENTRANCE & EU-commissioned Research project & 4 scenarios: Directed-transition, Techno Friendly, Societal Commitment, Gradual development & 3x1.5°C and 1x2°C & 2050 & EU + Norway and Turkey & 2020 & 3-dimensional scenario space; each dimension represents a key uncertainty: technological development, policy exertion, societal attitude \\[22pt]
            \cmidrule[0.5pt]{1-8}
            \gls{TYNDP} & ENTSOG {\&} ENTSOE incl. stakeholder process and public consultation (not funded project) & 2 scenarios + national trends based on NECPs & 1.5°C for 2 scenarios (Distributed Energy, Global Ambitions), not relevant for 3rd scenario - National Trends & 2050 & EU & 2022 (latest quantification), updated in 2024 (only qualitatively) & Objective is neutral perspective and widely accepted data and assumptions. The primary focus is the long-term development of energy infrastructure \\[22pt]
            \cmidrule[0.5pt]{1-8}
            \gls{WEC} & World Energy Council (stakeholder informed scenarios) & 2 scenarios: Rivers and Rocks & around 1.5°C and above 2°C respectively & 2060 & World and 8 major regions including Europe & 2024 & Fractured world but climate change remains a priority (focus is on what actions and when to act), with a shift of the economic growth to Asia\\[22pt]
            \cmidrule[0.5pt]{1-8}
            \gls{IRENA} & International Renewable Energy Agency & 2 scenarios: Planned Energy Scenario and 1.5°C scenario & above 2°C and around 1.5°C and respectively & 2050 & World and G-20 economies & 2023 & Three transition pillars: physical infrastructure, policy and regulation, skills and capacities of communities and individuals to drive and sustain the energy transition \\[22pt]
            \cmidrule[0.5pt]{1-8}
            Shell & Shell group & 2 scenarios: Archipelagos and Sky 2050 & above 2.2°C and around 1.5°C and respectively & 2100 & World and 16 regions, including international marine bunkers  & 2023 & Security is the major concern: energy security in Archipelagos and climate security in Sky 2050. Other factors include technology innovation, bottom-up societal changes, financing of the transition\\[22pt]
            \specialrule{2pt}{0pt}{0pt}
    \end{tabular}%
    }}
    \label{tab:literature1}
\end{table*}

\subsubsection{JRC-GECO Scenarios}
Complementing the \gls{EU} Reference and 2040 Climate Target Plan scenarios, the \gls{GECO} \cite{JRC136265} provides an updated overview of the impact of energy and climate policies worldwide on meeting the goals of the Paris Agreement. 

The scenarios in \gls{GECO} include the Reference scenario, which reflects how the global energy system and emissions will evolve based on existing policies and targets. Building on this, the Fossil Fuels Subsidies Phase-Out scenario assumes the complete elimination of fossil fuel subsidies from 2025 onwards. In this set, the NDC-LTS scenario is more ambitious, encompassing both the medium-term targets of the NDCs and the long-term targets from the LTSs, and accounting for decarbonization efforts in the international aviation and maritime sectors. The NDC-Only scenario isolates the impact of additional policies excluding the influence of the LTSs. Lastly, the 1.5°C scenario aims to limit the global temperature increase to 1.5°C over the century, constrained by 2020--2100 carbon budget. Likewise, 
JRC-LCEO (2018) \cite{JRC-LCEO} represents another key set of scenarios developed under European Commission studies, which has been comprehensively discussed in \cite{HAINSCH2022122067}.

\subsection{European Scenarios in Research Projects}
In addition to its own analyses, the European Commission interacts and discusses with the expertise of the European research community, supporting energy and climate-related research through competitively funded programs. The most important European scenario studies of recent years in this field are presented below.

\subsubsection{openENTRANCE Scenarios}
The openENTRANCE project introduces a novel storyline approach based on a three-dimensional scenario space, where each dimension represents a key uncertainty: technological development, policy exertion, and societal attitude \cite{openentrance1,HAINSCH2022122067}. The closer a scenario is to the coordinate origin, the more conservative it is, hence the 2°C scenario : 'Gradual Development´. The remaining three scenarios (Societal Commitment, Techno-friendly, Directed Transition) describe combinations of two out of three exposed uncertain dimensions and refer to the 1.5°C temperature increase target. This linkage of the four quantified openENTRANCE   scenarios to global climate targets is another novelty of the openENTRANCE scenario framework.

Combined with fully open datasets and the open-source energy system model GENeSYS-MOD \cite{GENeSYS-MOD}, the openENTRANCE project set new standards in scenario building and model-based scenario quantification for the EU30+ region (including Turkey) up to the year 2050. For further details, refer to \cite{OEWEBSITE}.

\subsubsection{SET-Nav Scenarios and other EU research projects}
The SET-Nav project employs a traditional two-dimensional scenario-building approach with axes of uncertainty: cooperation vs. entrenchment and decentralization vs. path dependency (centralization) \cite{PedroSETNav2018}. This leads to four scenario narratives illustrating different future pathways for energy transition: (i) Diversification - a decentralized and cooperative world with many new market entrants, characterized by digitalization and prosumers. (ii) Directed Vision - a cooperative scenario with strong EU guidance and path dependency, favoring large actors. (iii) Localisation - a less cooperative scenario focusing on local resources, with national strategies differing by country. (iv) National Champions - a scenario minimizing transition costs, giving a strong role to incumbent firms and utilities. The diversity of the four pathways enables the investigation of a large number of drivers and technologies (notably Energy Efficiency, Renewable Energy Sources, Nuclear, and Carbon Capture \& Storage) in all important sectors (building, transport, industry). All SET-Nav scenarios are targeting 85\%-95\% decarbonization in Europe by 2050.

Several other research projects have contributed to the development of European decarbonization scenarios. Notable among these are REEEM (2019) \cite{REEEM}, MEDEAS (2018) \cite{MEDEAS}, and SUSPLAN (2011) \cite{SUSPLAN}. For a comprehensive discussion of these studies, refer to \cite{HAINSCH2022122067}.

\subsection{Selected Global Energy Transition Scenarios}
Apart from the \gls{EU} scenarios, examining global scenarios is crucial for policymakers. It enables them to better anticipate and respond to international trends and challenges, ensuring that Europe's energy transition is aligned with broader global efforts to combat climate change. This understanding is essential for navigating geopolitical risks, integrating cutting-edge technologies into \gls{EU} energy systems, assessing the effectiveness of \gls{EU} policies in global climate mitigation, anticipating market shifts, and aligning \gls{EU} strategies with international efforts.

\begin{table*}[htb!]
    \centering
    \scriptsize
    \caption{Reviewed scenarios at a glance -- Part 2 of 3.}
    \makebox[\textwidth][c]{%
    \resizebox{1\textwidth}{!}{
    \begin{tabular}{@{}F{35pt}T{50pt}T{70pt}T{50pt}T{23pt}T{50pt}T{35pt}L{80pt}@{}}
    \specialrule{2pt}{0pt}{0pt}\\[-4pt]
            \textbf{Name} & \textbf{Conducted By} & \textbf{Scenarios} & \textbf{Temp. Rise} & \textbf{Horizon} & \textbf{Coverage} & \textbf{Dev. Date} & \textbf{Key Considerations}   \\[5pt]
            \cmidrule[1pt]{1-8}
            BP & BP group & 3 scenarios: Momentum, Rupture, Current trends & 1.5 and 2 & 2050 & World separated in 6 regions  & 2023 & -- \\[8pt]
            \cmidrule[0.5pt]{1-8}
            Total & Total Energies group & 3 scenarios: Accelerated, Net Zero, New Momentum & 1.7, 2.1, over 3  & 2050 & World separated in 3 regions  & 2023 & Energy efficiency, electrification, decarbonised electricity mix, flexibility needs\\[15pt]
            \cmidrule[0.5pt]{1-8}
            EDF & EDFs group & 1 scenario: Net Zero & 1.5  & 2050 & Europe with country granularity  & 2024 & -- \\[8pt]
            \cmidrule[0.5pt]{1-8}
            EC EU Reference & -- & 1 scenario: EU Reference Scenario 2020 & not relevant & 2050 & EU-27  & 2020 & --\\[22pt]
            \cmidrule[0.5pt]{1-8}
            NGFS & supported by the NGFS network & 7 scenarios: Delayed Transition, Fragmented World, Net Zero 2050, Below 2, Low Demand, NDCs, Current Policies & 1.4 to 2.9 & 2050 & World  & 2023 & Mapping out how economies might evolve under different assumptions\\[17pt]
            \cmidrule[0.5pt]{1-8}
            SET-Nav & EU-commissioned research project & 4 scenarios: Diversification, Directed Vision, Localisation, National Champions & not relevant; but 85-95\% decarbonisation in EU by 2050 & 2050 & EU & 2018 & 2-dimensional scenario space; axes of uncertainty: cooperation/entrancement versus decentralisation/path dependency (centralisation) \\[18pt]
            \specialrule{2pt}{0pt}{0pt}
    \end{tabular}%
    }}
    \label{tab:literature2}
\end{table*}

\subsubsection{Shared Socioeconomic Pathways (SSPs)}
\glspl{SSP} are climate change scenarios of projected socioeconomic global changes up to 2100, as defined in the \gls{IPCC} Sixth Assessment Report on Climate Change in 2021 \cite{IPCC21}. They are used to derive greenhouse gas emissions scenarios under different climate policies. The SSPs provide narratives describing alternative socioeconomic developments, with storylines that qualitatively describe the logic connecting elements of these narratives and how they are linked to challenges for mitigation and adaptation. Quantitatively, they provide data on national population, urbanization, and GDP (per capita). 
Refer to \cite{SSP2017a} for a description of the five SSPs.

\subsubsection{IEA Scenarios (World Energy Outlook 2023)}
The \gls{IEA} publishes the annual World Energy Outlook providing comprehensive analysis and strategic insights into the global energy system. The \gls{IEA} 2023 scenarios, presented in the World Energy Outlook 2023 (WEO 2023) \cite{WEO2023}, explore different pathways for the global energy future based on climate commitments, current policies, economic shifts, and energy use. These scenarios include:

\vspace{20pt}
\begin{itemize}
\setlength\itemsep{-2pt}
\item[(i)] Stated Policies Scenario (STEPS), based on current and announced energy policies.
\item[(ii)] Announced Pledges Scenario (APS), including additional climate commitments announced by governments.
\item[(iii)]Net Zero Emissions by 2050 Scenario (NZE), aiming to achieve net-zero emissions by 2050.
\end{itemize}
  The \gls{IEA} scenarios consider recent geopolitical tensions, energy crises, and evolving energy markets, assessing pathways to maintain alignment with the 1.5°C target despite rising energy demand.
  
  The  key uncertainties in the scenario development are related to governmental policies, technological development, global economic dynamics and demographic trends. \gls{IEA}’s methodology combines sophisticated economic and energy models with qualitative analyses to assess the impacts of different energy transition pathways. The \gls{IEA} scenarios cover a period up to 2050 and include detailed regional analyses for major global economies. 

\subsubsection{IRENA Scenarios (World Energy Transition Outlook)}

In 2023, the \gls{IRENA} published its latest World Energy Transition Outlook \cite{IRENA2023} that builds on two key scenarios to capture global progress towards the Paris Agreement climate change mitigation targets. The Planned Energy Scenario serves as a benchmark, providing a perspective of the energy system based on governments’ energy plans and other planned targets and policies in place at the time of the analysis (2022). The 1.5°C Scenario is a normative scenario meeting the Paris Agreement goal of limiting the global average temperature increase by the end of the century to well below two degrees from the pre-industrial level by prioritising readily available technology solutions that can be scaled up as necessary to meet the goal. The 1.5°C Scenario highlights for 2050, and at a global scale, a more than 90\% share in renewable energy for electricity in 2050, more than 80\% share of renewables in final energy consumption, electrification of energy demand that is more than 2 times higher than today, a significant role of green hydrogen and CO2 capture upscaling to 7 Gt CO2 in 2050 (of which more than half will come from negative emissions technologies).

\gls{IRENA}’s scenarios focus on three key energy transition pillars: physical infrastructure, policy and regulation, and skills and capacities of communities and individuals to drive and sustain the energy transition. Main assumptions relate to energy policies (in place and planned), available technologies, and economic and demographic trends (from detailed macro-economic modelling). The scenarios cover the major world economies (G-20) in detail, and they span to 2050. The transparency of the scenarios is ensured via documentation of the main assumptions and provision of detailed results.

\subsubsection{WEC Scenarios and its Update in 2024}
\gls{WEC}, a UN-accredited body with more than 3000 members in 90 countries, publishes its stakeholder-informed scenarios every three years. The scenario development process starts around 1.5 years before their publication, with several regional workshops, in which a scenario storyline refreshment is performed and important scenario drivers are identified. Stakeholder interviews and desktop research complement the process of developing the scenario storylines at global and regional scales. The quantification of scenarios from 2013 to 2019 was performed by the Paul Scherrer Institute (PSI), while in 2024, it was done by Enerdata.

In 2019, the three scenarios had musical names  \cite{WEC2019}: Modern Jazz (focusing on a market-led disruptive world); Unfinished Symphony (describing a policy-led world to address major environmental and climate change challenges); and Hard Rock (describing a fragmented world with low cooperation). These three scenarios were based on eight key drivers – four predetermined factors common to all scenarios and four critical uncertainties differentiated across scenarios. The predetermined factors include the slowing growth rate of the global population, the rise of new technologies, the appreciation of planetary boundaries, and the shift in economic power to Asia. The four critical uncertainties are identified as the pace of innovation and productivity gains, the international governance and geopolitical changes, the priority given to climate change and connected issues, and the policy tools in action. 

For the 2024 scenario foundations \cite{WEC2024}, \gls{WEC} moved the state of geopolitics from key uncertainty to a pre-determined factor common to all scenarios. This ruled out the cooperative globalised world assumed. Instead, a more fractured world was taken in all scenarios. Another major change was that the climate change priority stayed uncertain, but by acknowledging that, most people recognise the urgency, with the main differentiation being how it is expressed in the different scenarios – what actions are taken and when. A third major change was that the policy tools for actions critical uncertainty were expanded to include not only the “what tools” but also the “who” and “how”, including two major modes of cooperation which provide a thematic basis for differentiating the new scenarios: collaboration and alignment. The outcome of these changes is that instead of three, now two new scenarios were produced for 2024. The Rocks scenario describes a world of blocs of regions with leaky barriers in trade, in which the energy system has a long tail of fossil fuel use, and decarbonisation occurs in some of the blocs. The Rivers, which is a scenario of shifting alliances, and in which trade remains international in principle but with security carve-outs, while there is turbulent but swift fossil fuel substitution (electricity, hydrogen, biofuels) and cross-border connections enabled by technology.

\begin{table*}
    \centering
    \scriptsize
    \caption{Reviewed scenarios at a glance -- Part 3 of 3.}
    \makebox[\textwidth][c]{%
    \resizebox{1\textwidth}{!}{
    \begin{tabular}{@{}F{35pt}T{50pt}T{70pt}T{50pt}T{23pt}T{50pt}T{35pt}L{80pt}@{}}
    \specialrule{2pt}{0pt}{0pt}\\[-4pt]
            SUSPLAN & EU-commissioned research project & 4 scenarios: Green (High RES, centralised), Yellow (High RES, decentralised), Red (low RES), Blue (moderate RES) & not relevant & 2050 & EU & 2011 & 2-dimensional scenario space; axes of uncertainty: fast/slow technological development versus positive/indifferent public attitude \\[22pt]
            \cmidrule[0.5pt]{1-8}
            IEA-WEO2023 & IEA & 3 scenarios: Stated Policy Scenarios (STEPS), Announced Pledges Scenario (APS), Net Zero Emissions (NZE) & 1.5°C for 1 scenario: Net Zero Emissions (NZE) & 2050 & Global & 2023 & Main assumptions and key uncertainties in scenario development: governmental policies, technological development, global economic dynamics, and demographic trends \\[22pt]
            \cmidrule[0.5pt]{1-8}
            \glspl{SSP} & IPCC (Sixth Assessment Report) & 5 scenarios: Sustainability - Taking the Green Road (SSP1), Middle of the Road (SSP2), Regional Rivalry - A Rocky Road (SSP3), Inequality - A Road Divided (SSP4), Fossil fueled Development - Taking the Highway (SSP5) & 1.6°C-2.4°C (2041-2060); 1.4°C-4.4°C (2081-2100) & 2100 & Global & 2021 (latest update) & SSPs can be quantified with various Integrated Assessment Models (IAMs) to explore possible future pathways concerning \\[12pt]
            \cmidrule[0.5pt]{1-8}
            This study & EU-commissioned Research project& 3 scenarios + Extension of NECPs: EU Trinity, NECP Essentials, REPowerEU++, Go RES & 1.5°C/1.5°C/2°C for the three scenarios, not relevant for national trends & 2060 & EU& Developed in 2024 & Two scenarios are optimistic and pessimistic, respectively, toward the future of the European energy transition. One scenario seeks the independence of the EU energy system. To be quantified using GENeSYS-MOD. \\[22pt]
            \specialrule{2pt}{0pt}{0pt}
    \end{tabular}%
    }}
    \label{tab:literature3}
\end{table*}

\subsection{Selected Stakeholder Association and Energy Company Scenarios}

\subsubsection{ENTSOG's and ENTSOE's TYNDP 2022 Scenarios and its Update in 2024}

ENTSOG and ENTSOE, the European Network of Transmission System Operators for Gas and Electricity, aim to refrain with their biannual published Ten-Years Network Development Plans (TYNDP) from making political statements through their underlying scenarios and strive to ground key parameters in widely accepted data and assumptions. TYNDP scenarios, furthermore, are expected to maintain robustness by ensuring consistency between successive TYNDP publications. 
The TYNDP 2022 is based on extensive stakeholder engagements and a public consultation conducted in 2020 and presents three different long-term scenarios compliant with the Paris Agreement, starting from 2030 and reflecting increasing uncertainties towards 2050 \cite{TYNDP2022}: National Trends, Distributed Energy, Global Ambitions. The National Trends scenario operates within an input framework derived from official data sets like PRIMES and the official energy and climate policies of \gls{EU} Member States (\glspl{NECP}, hydrogen strategies, etc.). The COP21-compliant scenarios (Distributed Energy, Global Ambitions) allow for more innovation to achieve a more ambitious decarbonization of the energy system by 2050. However, ENTSOG and ENTSOE take care to maintain a neutral perspective on the input sets and do not intend to use these scenarios to advocate for one political agenda over another. The primary focus of the TYNDP scenarios is the long-term development of energy infrastructure, and the differences between the two COP21-compliant scenarios primarily relate to potential variations in demand and supply patterns. 

For the TYNDP2024 no quantitative scenario results are published yet, only the Methodology Report \cite{TYNDP2024} is available. The TYNDP 2024 scenario updates mainly incorporate feedback given by stakeholders since 2022 and some modelling innovations to better capture the dynamics of the fast-changing European energy system. For the three known scenarios (defined in TYNDP 2022) quantitative results will be delivered for different time horizons. National Trends+ scenarios will still be presented for the 2030 and 2040 time horizons, since no data is available for 2050 in \gls{EU} Member States. The Distributed Energy and Global Ambitions scenarios will deliver the time horizon 2040 and 2050 to better cover increased uncertainties. For the latter two scenarios, the National Trends+ 2030 settings are used as a starting point and then evolve for the 2040 and 2050 time horizons according to their individual narratives.

\subsubsection{CO2 Value Europe (CVE) Scenarios}

The scenarios developed by CO2 Value Europe (CVE) \cite{CVE} are based on a simplified and intuitive methodology, similar to several other scenario development practices, which reduce key factors to just two. In this effort, the primary factors considered are the ``Level of Economic Activity'' and the ``Level of Support for \gls{CCU}''. The selection of these two factors out of nine initial candidates was based on their impact versus their uncertainty. While the level of support for \gls{CCU} is undoubtedly a significant factor in shaping future energy systems, its inclusion as one of the primary factors could be seen as a preference rather than the direct result of the uncertainty versus impact analysis. Nonetheless,  several other significant factors, such as geopolitical instability, energy policies, societal attitudes, and more have been elaborated upon within the narrative of each scenario.

\subsubsection{NGFS Scenarios}
The Network for Greening the Financial System (NGFS) scenarios \cite{NGFS} were designed to support policymakers, financial institutions, and other public and private stakeholders to understand the potential impact, risks, and opportunities of climate change and the transition to a low-carbon economy. The NGFS scenarios are built upon an uncertainty matrix with two axes: (i) Physical risks, i.e. risks linked to climate change such as more frequent or severe weather events like floods, droughts, and storms as well as other risks coming from an increase in global temperature. (ii) Transition risks,  including policy and regulation, technology developments and consumer attitudes.

The seven NGFS scenarios in total are classified as follows: Three ``Orderly'' Scenarios (Net Zero 2050 (1.5°C), Low Demand, Below 2°C) assuming that climate policies are introduced early and become gradually stricter. One ``Disorderly'' Scenario (Delayed Transition) with delayed and diverged policies across countries, inducing a high transition risk. Two ``Hot House World'' Scenarios (NDC, Current Policies) with climate policies being implemented but remaining insufficient to significantly halt climate change. One ``Too Little Too Late'' Scenario (Fragmented World) with a late and uncoordinated transition failing to limit physical risks.

The NGFS scenarios are deeply described in \cite{NGFSSlides} and \cite{NGFSReport}. They were initially developed in 2020 and have been updated periodically, with the latest version released in 2023. They include the latest geopolitical developments and cover a global scope and extend to the year 2100. The scenarios consider various uncertainties and driving forces, including policy implementation timelines, technological advancements, and socio-economic changes.

The NGFS scenarios are developed with a high level of transparency, incorporating input from a consortium of renowned academic research institutions. Detailed documentation and data are publicly available, making the assumptions and methodologies clear and accessible. The NGFS scenarios are created using three categories of models: (i) Physical risk models for providing climate and economic indicators (ii) Three Integrated Assessment Models (REMIND-MagPIE, GCAM and MESSAGEix-GLOBIOM) focusing on the energy sector, emissions and land use. (iii) A macro-financial model (NiGEM) for understanding impacts on key macro-financial fundamentals.

\subsubsection{Selected Energy Company Scenarios}
Some global acting energy companies publish their own energy outlooks and transition scenarios. The methodologies used in these scenarios involve detailed modelling and analysis of energy demand and supply, incorporating various uncertainties and driving forces. This also includes assumptions about technological advancements, policy developments, geopolitics, and economic growth. However, only reports and presentations are available while methods, modelling approaches, and data remain confidential. None of these scenarios of energy companies are EU-commissioned. As part of this literature review, these scenarios give another perspective to the broader debate on energy transition and give insights to strategic planning of key energy companies.

BP’s Energy Outlook 2023 presents three main global scenarios \cite{BP}: Accelerated, Net Zero, and New Momentum. The Accelerated scenario envisions rapid policy and technological advancements leading to significant reductions in carbon emissions. The Net Zero scenario aims to balance emissions produced and removed by 2050, relying heavily on renewable energy and carbon capture technologies. The New Momentum scenario reflects current policy settings and technological trends, resulting in slower progress towards decarbonization. 

Shell has also been developing energy scenarios for more than 50 years. In the latest Shell’s scenario edition \cite{Shell2024}, the recent developments in the energy sector triggered by the COVID pandemic and the Russian-Ukraine crisis are considered, leading to two scenarios in which the security mindset dominates and national interest takes precedence over the political agenda. Energy prices, supply concerns, and climate pressure are elements used in the scenario storylines. The Archipelagos scenario is an exploitative one, seeking to follow a possible path from where the world was in 2022. In this scenario, the energy security mindset that is dominant today becomes entrenched worldwide while aiming at a global average surface temperature increase of around 2.2°C with increasing use of low-carbon technologies and deployment of hydrogen mainly after 2050. The Sky 2050 takes a normative approach that starts with a desired outcome and works backwards to explore how that outcome could be achieved. In Sky 2050, the world aims to achieve two key objectives: net-zero emissions by 2050 and a limit on the average global temperature increase to well below two degrees from the pre-industrial levels by the end of the century. In Sky 2050, long-term climate security is the primary anchor, in contrast to Archipelagos, where the emphasis is on energy security. Sky 2050 achieves net-zero emissions rapidly by retiring fossil-fuel assets, imposing punitive carbon prices, scaling early-stage technologies, and conserving energy through efficiency and austerity. Societal choices are restricted, such as limiting internal combustion engine vehicle sales, meat consumption, and opposition to wind turbine and power line construction in unwelcome areas.   

TotalEnergies’ Energy Outlook 2023 outlines three scenarios \cite{Total} Momentum, Rupture, and Current Trends. The Momentum scenario assumes a continuation of current policies and trends, leading to moderate decarbonization. The Rupture scenario envisions a radical shift towards sustainability, driven by aggressive policy measures and technological breakthroughs. The Current Trends scenario projects the future based on existing policies without significant changes. Key features include the role of energy efficiency, the impact of regional differences, and the importance of renewable energy sources. The results suggest that the Rupture scenario could lead to a significant reduction in global carbon emissions, but it requires substantial policy intervention and technological innovation. These scenarios have a global spatial scope, divided into main regions: NZ50 (countries engaged in carbon neutrality in 2050), China and ``Global South''.

EDF’s Net Zero 2050 scenario \cite{EDF} focuses on achieving European carbon neutrality by mid-century by reducing energy demand and minimizing the use of fossil fuels. Three main levers are included: electrification of end-uses, energy efficiency and behavioural changes. Key features include assumptions about the growing role of electricity in the energy mix, via mass electrification of light transport and buildings; a limited supply of bio-energies which should thus be reserved for uses for which no decarbonized alternatives exist; and the necessity to target the use of decarbonized molecules given their high costs and energy requirements. The results highlight that achieving Net Zero by 2050 is feasible but requires a comprehensive approach, including significant investments in a diversified electricity mix of renewable energies including hydropower, nuclear and decarbonized thermal generation, major changes to the electricity transmission and distribution networks, and greater flexibility and storage, both on generation and demand side. 

\subsection{Solely Qualitative Scenarios}

In the literature, there is a great variety of energy-related qualitative scenarios, for example focusing on future developments of different primary fuels and sectors, such as gas (e.g. \cite{WeissPoganietzPoncette2021}), oil (e.g. \cite{Mottaghi2019, Norouzi2019}), transport (e.g. \cite{VenturiniHansenAndersen2019}), heating (e.g \cite{ZivkovicEtAl2016}) or electricity (e.g. \cite{MierEtAl2023}. These scenarios also have a focus on specific aspects such as energy security (e.g \cite{Mansson2016}) or energy demand (e.g. \cite{EhrenbergerEtAl2021}) and are global (e.g. \cite{AnsariHolz2019}), European (e.g. \cite{CaprosEtAl2014, ShoaiTehraniDaCosta2013}) or national (e.g. \cite{hoffart2021_1, VenturiniHansenAndersen2019} in scope. Different methods, such as Cross-Impact Balance (CIB) (e.g. \cite{KunzVoegele2017}), morphological (e.g.\cite{PereverzaEtAl2017}) or formalized approaches (e.g. \cite{Hoffart2022}) as well as hybrid studies combining qualitative and quantitative scenario approaches (e.g \cite{VenturiniHansenAndersen2019, AnsariHolz2019, McDowall2014}) are used. As an example for European scenarios, Schanes et al. \cite{SchanesJaegerDrummond2019} developed three scenarios: (i) In the ``global cooperation'' scenario, the world unites in coordinated efforts to reduce resource use through multilateral actions. (ii) In ``Europe goes alone'' Europe independently spearheads the transition to a resource-efficient economy with strong regional leadership. (iii) The ``Civil society leads'' scenario envisions grassroots initiatives in Europe, where citizens and local communities drive resource efficiency and sufficiency.

\section{EU EnVis-2060: Qualitative Element} \label{sec:qualitative}

Scenario definition often begins by identifying a focal issue or decision \cite{chermack2011a}. In this context, the focal issue is the transition of the European energy system. As evident from the literature review in the previous section, the issue of long-term energy transition is not a new problem but rather a well-established one with numerous existing methods, approaches, insights and underlying qualitative and quantitative scenarios. Nevertheless, the development of new scenarios remains crucial due to the pressing significance of this issue and the rapidly evolving global context. The scenarios developed in this paper aim to address the European energy transition by incorporating recent policies and initiatives from the European Commission, with a specific emphasis on the \emph{\glspl{NECP}} \cite{NECP}, \emph{REPowerEU} \cite{REPowerEU}, and the \emph{Fit for 55 package} \cite{Fitfor55}.

This set includes three scenarios, each with distinct features exploring different aspects of the energy transition. Additionally, one scenario replicates \glspl{NECP} \gls{WEM} until 2040 and extends it towards 2060, incorporating minimal changes in the policies and trends outlined in these plans. These scenarios are designed to provide insights into the potential outcomes of current and future energy policies, identifying key drivers, opportunities, and challenges. They project into the future until 2060 and cover European countries geographically. 

For clarity and ease of reference, the scenarios developed in this study are named \textbf{``\gls{EnVis}''}, encapsulating the essence of our study.

\subsection{Workflow}

The qualitative scenarios in this study were primarily developed by the core team (the authors of this paper) and reviewed by an extended team, as acknowledged in the acknowledgements section. The core team comprised six groups, each consisting of one to four members. The process of developing the initial version of qualitative scenarios spanned approximately six months and included the following steps:
\begin{enumerate}[leftmargin=*]
\setlength\itemsep{-2pt}
\item In two kick-off workshops, the core team engaged in thorough discussions to establish the methodology and brainstorm key driving forces influencing the energy transition. Parallel meetings were also held solely to discuss the methodology. Once the methodology was chosen, all partners were asked to propose concepts for three to four scenarios in a three-dimensional space, with each dimension representing a primary key factor.
\item All partners proposed their ideas for the three primary uncertainties. These proposals were discussed in another meeting, and the three primary uncertainties were selected based on consensus and voting.
\item The three primary uncertainties, along with the employed methodology, were presented to the extended team---relevant stakeholders, including need owners and collaborators within a broader scope---in a workshop for their input.
\item Based on the three finalized uncertainties, each group proposed three to four scenarios. These ideas were then presented ideas in a workshop, followed by a discussion after each presentation.
\item The proposed ideas were merged into four scenarios. These merged scenarios were reviewed individually by each team and by the extended team. Inputs were discussed with the core team, and the narratives and focus areas were finalized accordingly.
\item Additional key uncertainties, initially excluded, were incorporated into the qualitative scenarios through a morphological box.
\item Based on the status of key uncertainties in each scenario, relevant indicators and factors for energy modelling tools were detailed in a \acrfull{Q2Q} matrix. This matrix was reviewed by both the core and extended teams for further inputs.
\item The final element of the qualitative scenarios, the \gls{Q2Q} matrix, was translated into quantitative scenarios (version-00). The preliminary outcomes for two of the scenarios are presented in Section \ref{sec:quantitave}. These results will provide feedback for refining the qualitative scenarios, with modifications expected to reach the ultimate set of scenarios.
\end{enumerate}

During this period, we conducted six two-hour workshops with most core team members, supplemented by additional meetings within specific teams and three presentations to the extended group. Between workshops, specific tasks---such as proposing scenario narratives, as explained further in the next sections---were assigned to the teams, with the workshops serving as forums for discussing and refining the results of these tasks.

\subsection{Methodology and Results}

\begin{figure*}
    \centering
    \makebox[\textwidth][c]{%
    \includegraphics[width=1.12\textwidth, trim=0pt 0pt 0pt 0pt]{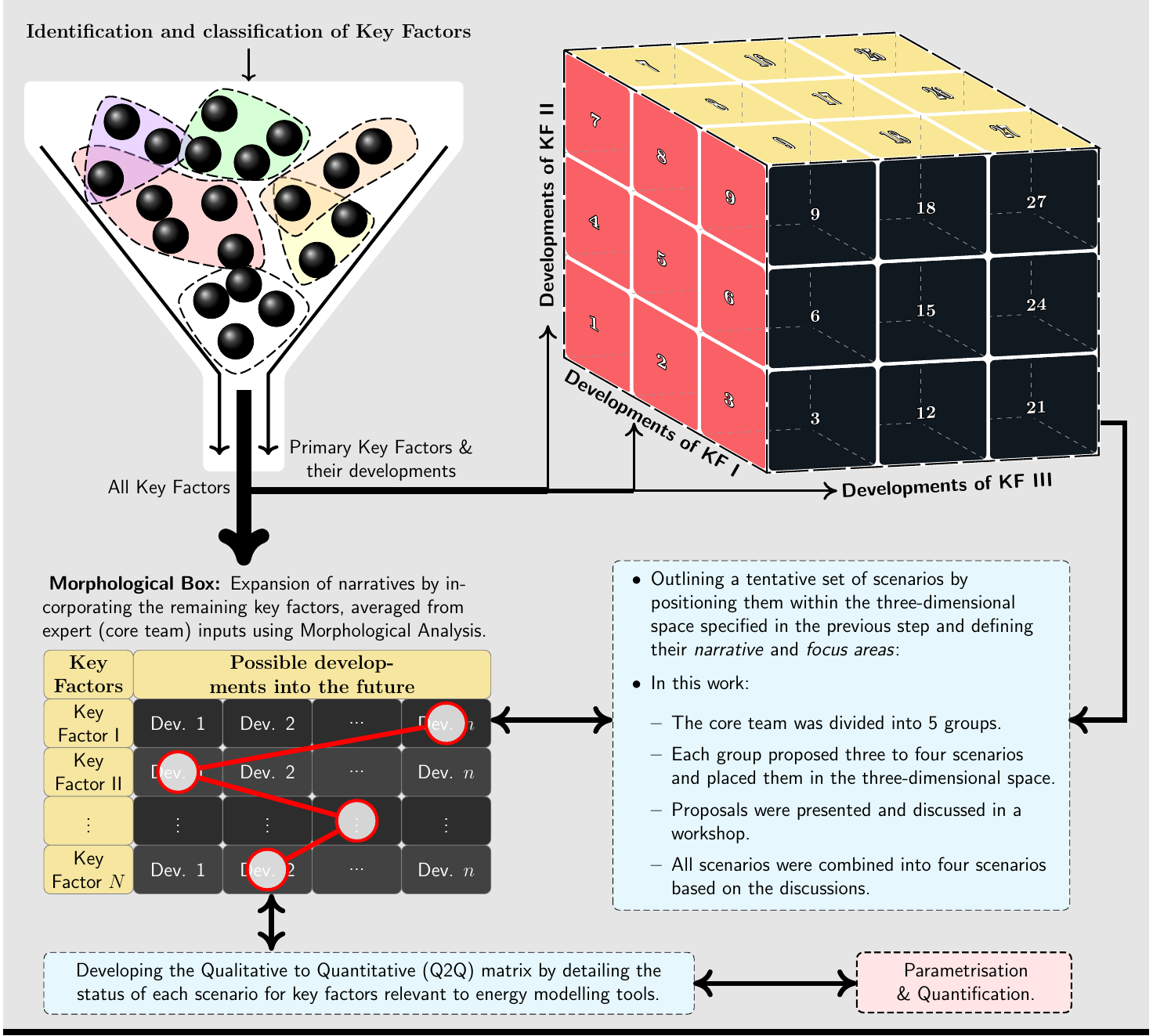}}
    \caption{The overall steps of the technique used in this study.}
    \label{fig:methodology}
\end{figure*}

While this study utilizes a combination of creative-narrative and formalized methods, it predominantly favours a creative-narrative approach for three main reasons. First, the core team's expertise aligns well with this methodology. 
Second, it provides flexibility for long-term investment planning, making it easier to incorporate explorative elements related to overall project goals. 
Lastly, this approach has a long-standing history of use in energy transition scenario development.
Figure \ref{fig:methodology} illustrates the overall steps of the technique used in this study.

\subsubsection{Step1: Exploring and classifying the key uncertainties} \label{sec:key_factors}

This section delineates the key factors influencing the long-term energy transition. These factors encompass both uncertainties and determinants. The first step is identifying the key uncertainties through literature review and brainstorming. The brainstorming, in this study, was conducted in different sessions with both core and the extended teams both in an unorganized manner by discussions and an organized manner by asking specific questions regarding the factors influencing various elements of an energy transition. For instance, some of these questions were the following:

\begin{itemize}[leftmargin=*]
    \setlength\itemsep{-4pt}
    \item \textcolor{black!70}{\textsl{What key factors influence the long-term demand projections in Europe (electricity, heat, and transportation)? See Figure \ref{fig:brainstorming} in the appendix.}}
    \item \textcolor{black!70}{\textsl{What key factors influence the future technology investment/operational costs in the European energy systems?}}
    \item \textcolor{black!70}{\textsl{What key factors influence the availability of import/exports related to energy systems from/to Europe in the future?}}
\end{itemize}

\vspace{20pt}
After identifying the key factors through brainstorming and literature review, these were classified into specific groups based on \acrshort{STEP} analysis — \textbf{S}ocial, \textbf{T}echnological, \textbf{E}conomic, and \textbf{P}olitical. The well-known \acrshort{STEP} analysis is a valuable tool for structuring thought processes and ensuring that no critical categories are overlooked.  It offers a logical and effective approach to examining and systematically exploring the external environment. \cite{chermack2011a}. Sometimes, another E, standing for Environmental, is added, becoming \acrshort{STEEP}, which is not the case in this study. Geopolitical instability can be considered within these groups, for example within the political aspect. However, due to its significant role in today's European energy transition, highlighted during the Russian invasion of Ukraine, we considered it as a separate group added to \acrshort{STEP} to form \acrshort{STEP+G}, i.e., \acrshort{STEP} plus Geopolitics. 

For energy transition, \acrshort{STEP} can be analyzed from two distinct perspectives: how these aspects influence long-term energy transition and vice versa. For example, regarding social implications, one perspective examines how the energy transition impacts societies. Transitioning to renewable energy can improve air quality and reduce health risks like respiratory and skin diseases by reducing pollution from fossil fuel use \cite{Ana2023a}. The perspective of this study, however, considers how society influences the energy transition.

\vspace{10pt}\noindent\textbullet \textsl{  Social Dynamics Toward Transformation:}\vspace{4pt}

Achieving net zero \gls{GHG} emissions is for \emph{people}, about \emph{people}, and done by \emph{people}. A well-informed society is essential for achieving decarbonization goals. This dimension has been acknowledged in numerous energy transition scenarios from various aspects (e.g., \cite{openentrance1, SSP2017a, susplan2011a}) and significantly influences the trajectory of the energy transition. The attitudes, behaviours, and societal shifts related to energy transition play a critical role in determining the feasibility and direction of \gls{GHG} emission reduction objectives. Societal awareness of environmental concerns is arguably the most important social factor influencing the energy transition. With heightened awareness, society can accept and support new technologies such as heat pumps, policies such as carbon pricing, and innovative programs like direct load control in the power sector \cite{barani2024a}, all of which facilitate achieving \gls{GHG} reduction targets. Behavioural shifts in consumption patterns, both in products and energy use, also require a well-informed society. Ultimately, a society highly aware of the consequences of global warming will demand environmental action. Recognizing the impact of social dynamics is crucial in shaping energy transition scenarios. In the \textbf{EU EnVis-2060} scenarios, ``Social Dynamics Toward Transformation'' is identified as a primary key factor.

\vspace{10pt}\noindent\textbullet \textsl{ Innovation:}\vspace{4pt}

Innovation is the \textit{compass} guiding us through the \textit{uncharted waters of energy transition}, where each \textit{innovation} is a step towards sustainability. It serves as a critical driver and, simultaneously, a source of uncertainty in the energy transition across various sectors such as power, heat, and transportation. Technological advancements, such as \glspl{RER}, \gls{ESS}, \gls{CCS}, green hydrogen, advanced nuclear reactors, biotechnology, and material science, as well as disruptive innovations, have the potential to reshape the pathways toward decarbonizing the energy system. According to the \gls{IEA}, the most significant innovation opportunities lie in advanced batteries, hydrogen electrolysers, and \gls{DACS} \cite{IEA2021a}. These three innovations alone constitute the primary differences between most long-term European energy transition scenarios.

Furthermore, energy efficiency-related innovations, encompassing advancements in energy management systems, smart appliances, building materials, industrial processes, and more efficient technologies such as heat pumps, are integral components of this transition.

Additionally, policy and planning innovation, exemplified by initiatives like Horizon 2020, play a significant role in the energy transition. The ongoing trends of decentralization (e.g., increasing integration of distributed generations \cite{barani2018a} in the power sector) and digitalization (e.g., smart grids \cite{barani2021a} in the power sector) also influence the energy transition. Moreover, financial innovations, such as green bonds, power purchase agreements (PPAs), and hydrogen purchase agreements (HPAs), mobilize capital toward renewable energy projects, opening new avenues for investment and propelling the transition toward a low-carbon future. 

Collectively, these key factors underscore the transformative potential of innovation in driving the decarbonization of the energy system while acknowledging the uncertainties and complexities it introduces.

\vspace{10pt}\noindent\textbullet \textsl{ Geopolitical Instability:}\vspace{4pt}

Geopolitical instability, characterized by uncertainty and turbulence arising from interactions between different countries, influences the energy transition in various ways. One key aspect of this is the geopolitics of energy. Since the Industrial Revolution, the political and strategic control and distribution of energy resources have played a crucial role in driving global prosperity\cite{pascual2010a}.

For instance, an article published about a decade and a half ago emphasized the geopolitical tensions created by fossil fuel scarcity and the potential of peak oil supply \cite{Bradshaw2009a, Hirsch2005a}. A little more than a decade later, another study described the world as experiencing ``relative fossil fuel abundance'' \cite{Mathieu2021a}. However, recent events, such as the Russian invasion of Ukraine, have brought resource scarcity back into focus.

\vspace{10pt}\noindent\textbullet \textsl{ Political will, Policies, and Regulations:}\vspace{4pt}

At the heart of the energy transition lies political will, which is essential for crafting the necessary policies, regulations, and legislation. These policies and regulations will profoundly influence technological advancements, economic strategies, and societal priorities, while concurrently being shaped by these evolving factors. To illustrate, the European Climate Law (\cite{EUClimateLaw2021}) has the most significant impact on the European energy transition in an energy modelling tool, surpassing any other factors discussed in this study. It mandates a reduction of net greenhouse gas emissions (emissions after deduction of removals) by at least 55\% compared to 1990 levels by 2030 and aims for no net emissions of greenhouse gases by 2050.

The effectiveness of Europe's progress towards these binding targets is supported by subsequent policies and regulations, such as carbon pricing, market designs, \gls{ETS}, subsidies and incentives supporting \gls{RD} in clean technologies and reduction of \gls{GHG} emissions. Additionally, lobbying plays a significant role, as various interest groups advocate for policies that align with their objectives, thereby impacting the legislative process.

\vspace{10pt}\noindent\textbullet \textsl{ Economic Factors:}\vspace{4pt}

The economy plays a crucial role in influencing the energy transition. A powerful economy is needed to provide the capital to enhance the infrastructure and replace existing GHG-emitting technologies with clean technologies in all sectors. Economic factors are also important to provide funding for \gls{RD} to develop innovative technologies that reduce \gls{GHG} emissions, improve energy efficiency, and shift industries away from fossil fuels. A powerful economy can also subsidize and incentivize the efforts to reduce \gls{GHG}. On the other hand, economic policies aimed at \gls{GHG} reduction, such as increasing carbon prices, can pose challenges to industrial and economic growth by raising operational costs for businesses and potentially slowing down production \cite{Boonman2024a}. In this regard, the European Green Deal has been designed as the EU’s strategic plan to transform the EU into a clean, resource-efficient, and competitive economy.

\vspace{10pt}\noindent\textbullet \textsl{Population and other relevant Factors:}\vspace{4pt}

Population influences the long-term energy transition by affecting energy demand in all sectors. In this study, we assume the same demographic changes, including birth rates and immigration, across all scenarios based on current projections. To this end, the differences in the results can be attributed to other varying factors between the scenarios, facilitating a clearer comparison. Additionally, the uncertainty in demographic changes is relatively low, particularly in the short term.

Beyond the key factors discussed, the energy transition is also influenced by other factors, such as the acceleration of global warming, environmental degradation or burdens, and pandemic impacts, which are not directly considered in this study.

Note that all the aspects discussed in this section are interrelated. For instance, public attitudes and awareness can be significantly influenced by policies. Examples of such policies include incorporating relevant education into school curricula, implementing labelling schemes for energy-efficient products, and conducting public awareness campaigns \cite{refscen2013}. To ensure the plausibility and consistency of the scenarios, a correlation matrix can be employed to comprehensively assess these interrelations. For example, given the significant impact of policies and regulations on society, it is improbable to have an unaware public when such policies and regulations are strongly oriented toward a sustainable energy transition.

\subsubsection{Step 2: Selection of three key uncertainties and drafting the initial scenario narratives and focus areas}

In the second step, the initial focus was limited to three key uncertainties to develop the narratives and focus areas of the scenarios. The limitation was intended to limit the state space of the possible scenarios, facilitating the group discussions to reach a consensus on the overall picture of the scenarios. In this regard, the first three uncertainties discussed in section \ref{sec:key_factors}, namely: \textit{`Social dynamics toward transformation'}, \textit{`innovation'}, and \textit{`geopolitical instability'} were considered as the primary key factors. 

For each dimension, three potential future developments were identified (see Figure \ref{fig:Morphologic Box}). Considering these developments, their combinations yielded 27 potential scenarios. Each team independently developed its own set of scenarios and focus areas, selecting three to four scenarios from the available options. These proposals were subsequently merged to create the initial draft of the qualitative scenarios. During the merging process and the selection of the primary key factors, several considerations were taken into account: critical aspects of today's European energy system,  the plausibility of the developments of the identified uncertainties in relation to each other, and ensuring that the new scenarios were distinct from pre-existing ones. Figure \ref{fig:sample_merging} in the appendix illustrates the merging process for one of the scenarios. Notably, the results included both a pessimistic and an optimistic contrasting scenario, which emerged unintentionally but intriguingly. The resulting scenarios include:

\begin{enumerate}
    \itemsep0em 
    \item \textit{Current Trends}: Continuation of current trends without major change \textcolor{black}{(\textbf{\textit{NECP Essentials}})}
    \item \textit{Pessimistic Scenario}: Challenges or negative developments of the key factors (\textbf{\textit{EU Trinity}})
    \item \textit{Optimistic Scenario}: Positive developments of the key factors (\textbf{\textit{Go RES}})
    \item \textcolor{black}{\textit{Paradigm Shift Scenario}:} A partially explorative scenario, i.e., how to reach energy independence in the EU \textcolor{black}{(\textbf{\textit{REPowerEU\textcolor{red}{\textbf{++}}}})}
\end{enumerate}

\begin{figure*}
    \centering
    \makebox[\textwidth][c]{%
    \includegraphics[width=1.15\textwidth, page=1]{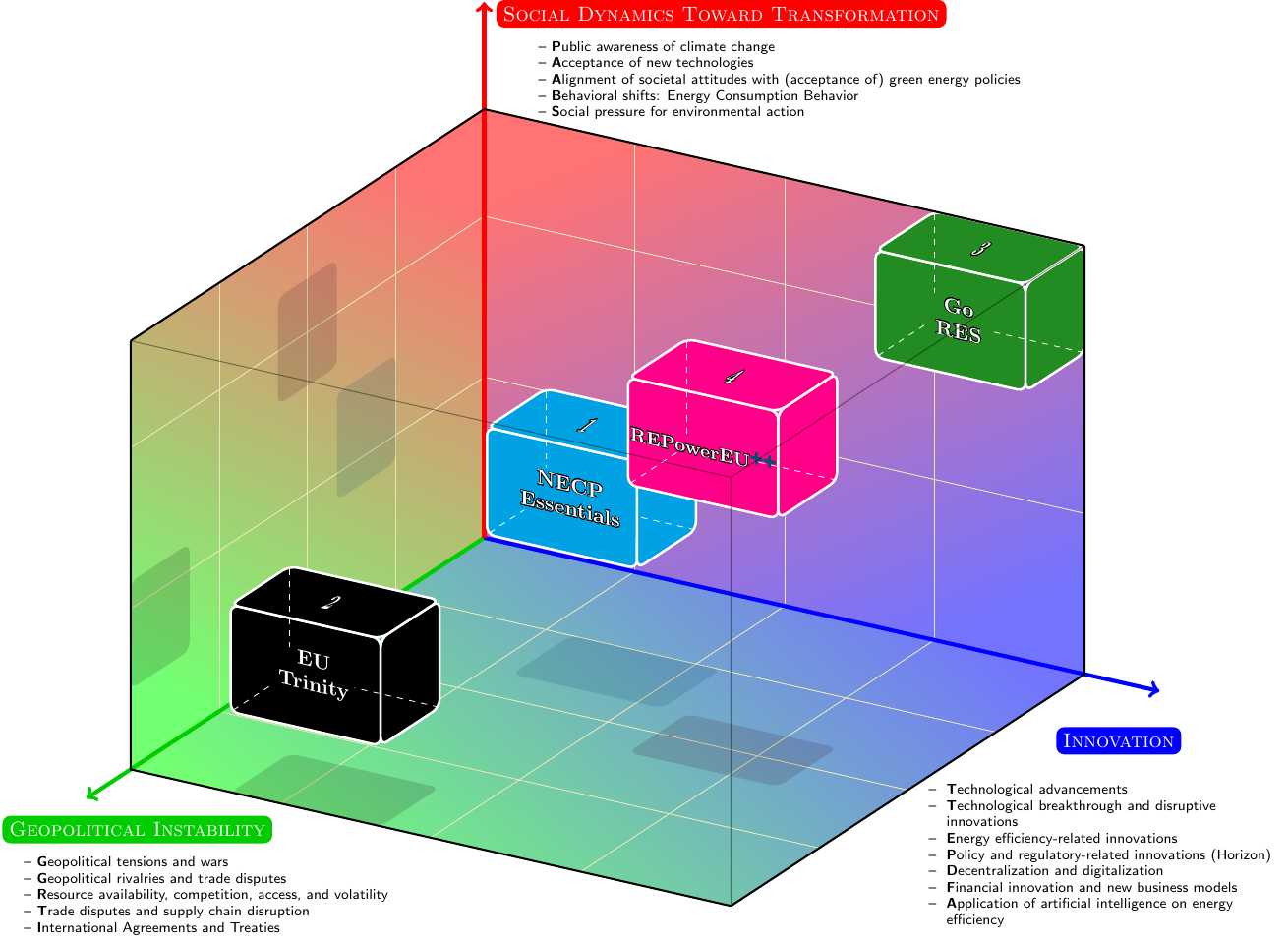}}
    \caption{The three-dimensional future space defined by \textit{Society}, \textit{Geopolitics}, and \textit{Innovation}, showing the positioning of the scenarios w.r.t these key factors.}
    \label{fig:3d_representation}
\end{figure*}

\begin{figure*}
\centering
\makebox[\textwidth][c]{%
\includegraphics[page=1]{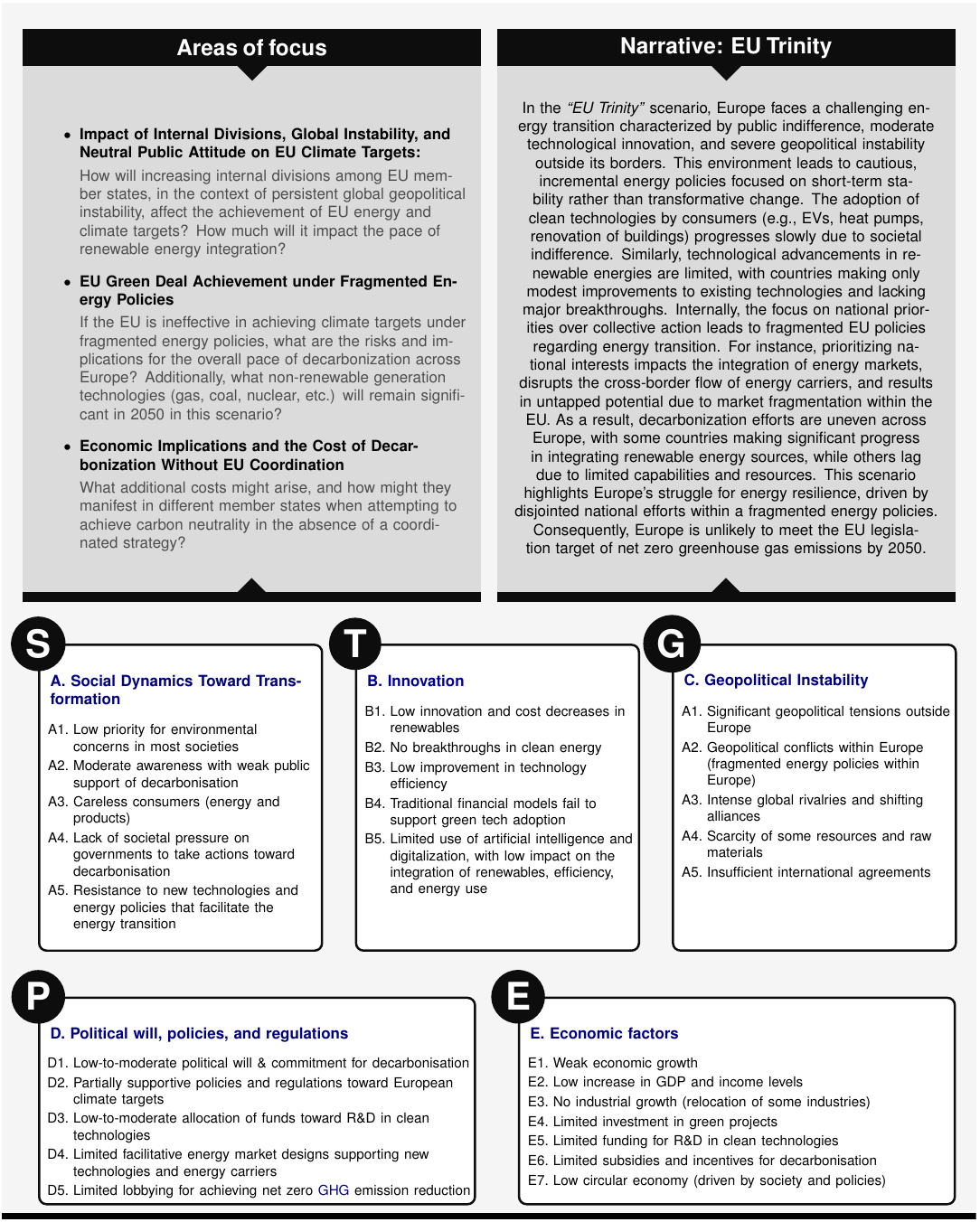}}
\caption{\textit{EU Trinity}: Narrative and focus areas.}
    \label{fig:narratives1}
\end{figure*}

\begin{figure*}
\centering
\makebox[\textwidth][c]{%
\includegraphics[page=2]{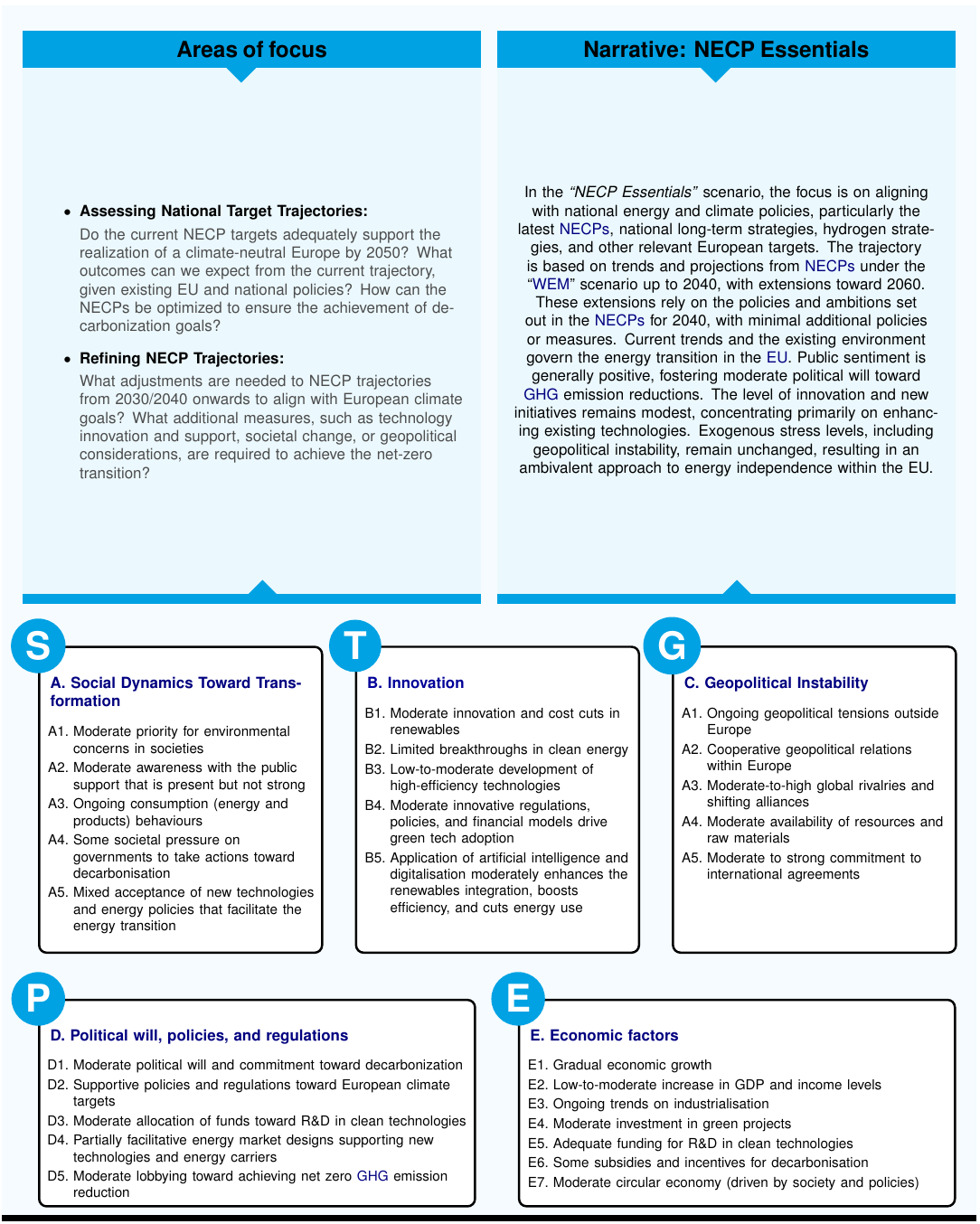}}
    \caption{\textit{NECP Essentials}: Narrative and focus areas.}
    \label{fig:narratives2}
\end{figure*}

\begin{figure*}
\centering
\makebox[\textwidth][c]{%
\includegraphics[page=3]{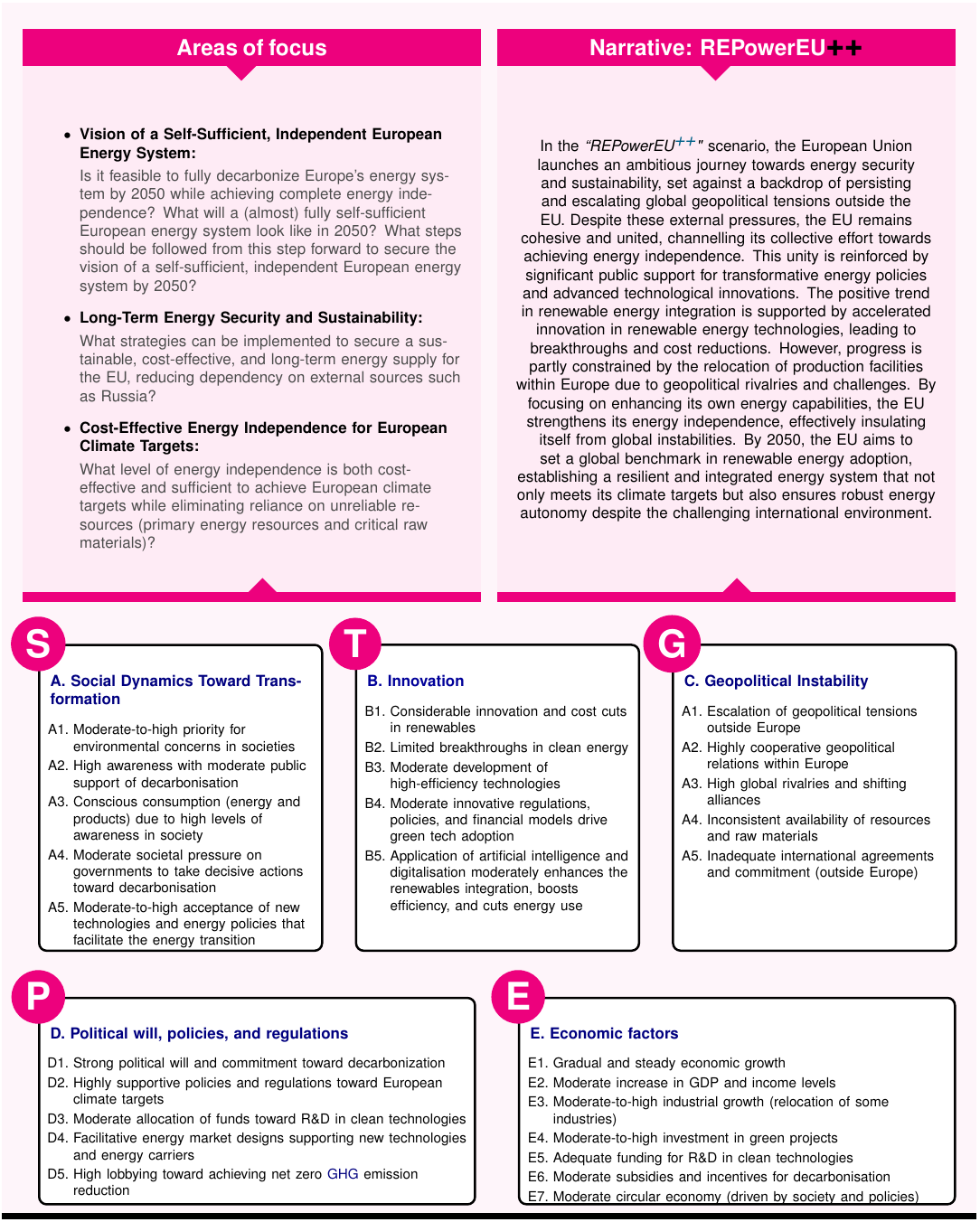}}
    \caption{\textit{REPowerEU++}: Narrative and focus areas.}
    \label{fig:narratives3}
\end{figure*}

\begin{figure*}
\centering
\makebox[\textwidth][c]{%
\includegraphics[page=4]{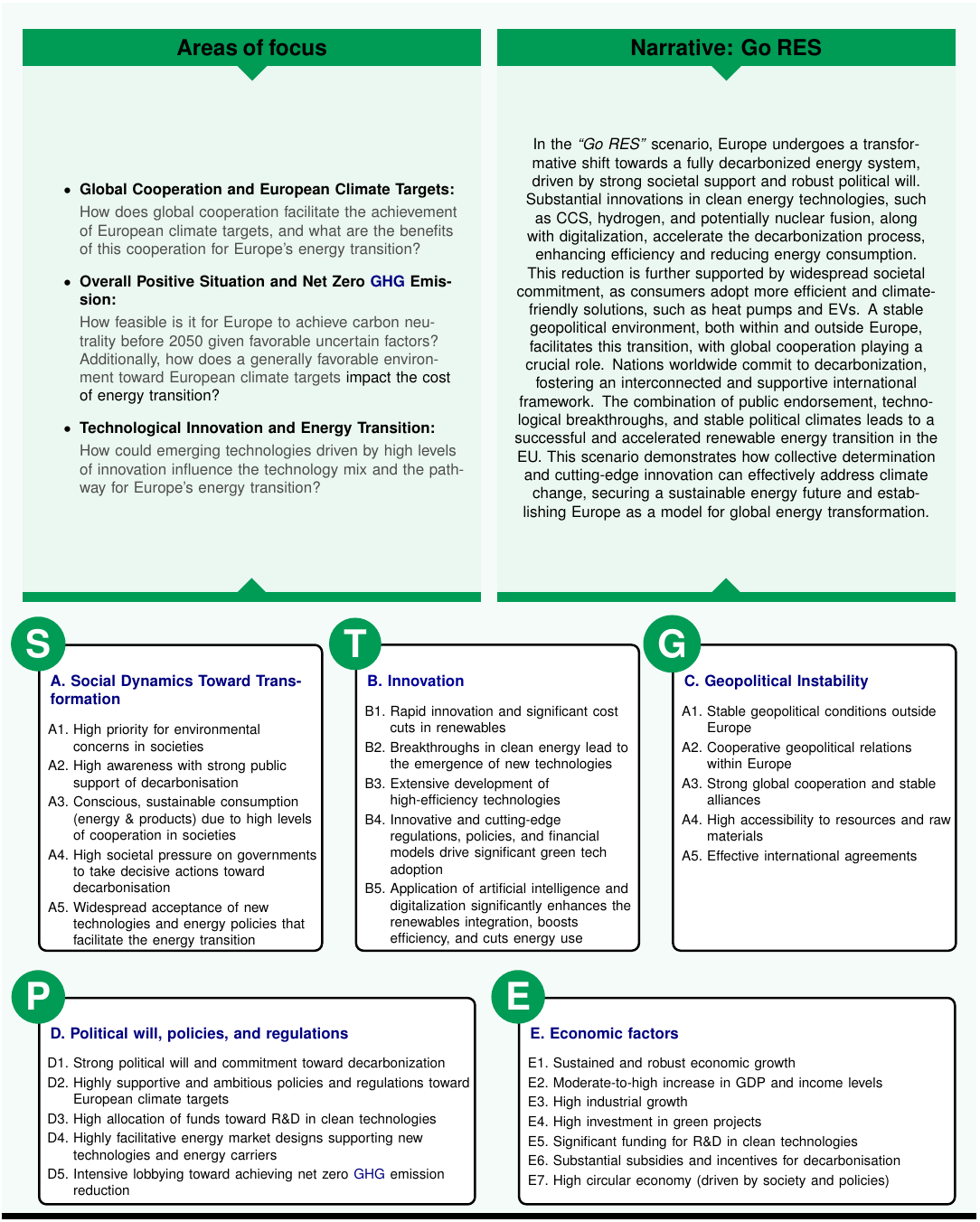}}
    \caption{\textit{Go RES}: Narrative and focus areas.}
    \label{fig:narratives4}
\end{figure*}

Figure \ref{fig:3d_representation} illustrates the positioning of these scenarios within the three-dimensional future space defined by the primary key factors. Based on the position and status of these key factors, corresponding narratives and focus areas have been delineated, effectively \emph{bringing the scenarios to life}. The inclusion of focus areas alongside the narratives serves to highlight the central themes of each scenario, as well as the specific questions that these scenarios aim to address once quantified. Figure \ref{fig:narratives1}, \ref{fig:narratives2}, \ref{fig:narratives3}, and \ref{fig:narratives4} present the narratives and their focus areas of the \gls{EnVis} scenarios.

\newgeometry{top=2cm, bottom=2cm, left=1cm, right=1cm}

\restoregeometry

\clearpage
\subsubsection{Step 3: Inclusion of Other Key Factors}

For energy transition scenarios, when the future state space is reduced to only two or three dimensions with two to three possible developments for each \cite{SSP2017a,openentrance1,SETNAV0,susplan2011a}, the other driving forces are largely neglected in the narrative development. Consequently, these narratives lack information about other aspects and driving forces. However, when these qualitative scenarios are planned to be quantified, other aspects must be considered, as quantification requires clear projections of all relevant factors. One approach is to make similar assumptions for the remaining driving forces and uncertainties. This can be useful if the goal is to explore the effects of the primary uncertainties considered in the first place, as the other influencing factors are similar in the scenarios. This is particularly valuable when the focus is on one or two factors.
However, two points must be considered: First, there are correlations between different uncertainty factors affecting the energy transition. For example, societal impact on political will. Assume that society is a primary key factor but political will is not (like in this study). If societal demand for \gls{GHG} emission reduction is extremely high in one scenario and low in another, the assumptions regarding the policies cannot be considered similar. Therefore, when the goal is to quantify these scenarios, consistency in assumptions is crucial. Second, these assumptions should be clearly stated in the qualitative scenarios for transparency.

\begin{figure*}[!htb]
    \centering
    \makebox[\textwidth][c]{%
    \includegraphics[width=1.1\textwidth, page=1]{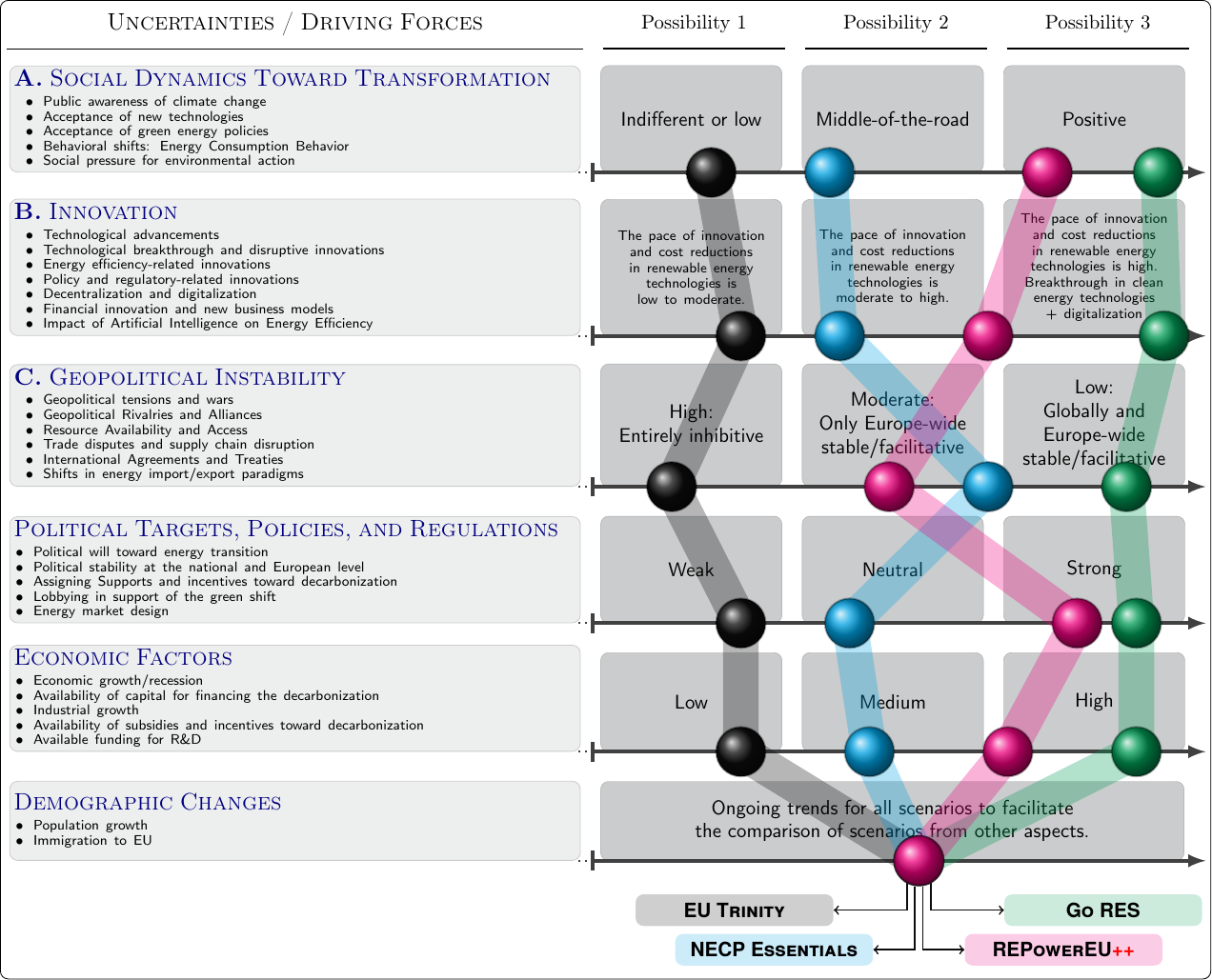}}
    \caption{Morphological Box: This figure summarizes the uncertainties considered in this study. The spheres represent the status of each key factor within the different scenarios of the \gls{EnVis} set.}
    \label{fig:Morphologic Box}
\end{figure*}

In this study, the development of the other key aspects is incorporated through morphological analysis. This process involves averaging the proposals from different groups within the core team, followed by discussions for calibration and consistency checks of the scenarios. The results are shown in Figure \ref{fig:Morphologic Box}. These additional aspects can also be included in the narratives, though they have not been added in this work. 
This approach could be employed for all driving forces at once without the three-dimensional space. However, achieving consensus on the primary focus of each scenario becomes challenging when the state space is extensive and the core team comprises many members.

\subsubsection{Step 4: Q2Q Matrix}

Up until this section, the driving forces were intentionally discussed in general terms to provide an overarching overview of the scenarios. For example, the innovation driving force was considered broadly, influencing various factors such as the projection of technology costs or the emergence of new technologies. This overview is then expanded upon in a detailed matrix called the \acrlong{Q2Q} (\gls{Q2Q}) matrix, which includes specific elements of the energy system. For instance, a narrative might describe innovation as moderate technological development. This aspect is further detailed in the \gls{Q2Q} matrix with projections on the availability and costs of various technologies such as onshore wind turbines. The main advantage of the \gls{Q2Q} matrix is its ability to facilitate and systematize the quantification process. Additionally, this comprehensive matrix enhances the clarity and readability of the qualitative scenarios. Table \ref{tab:Q2Q} (Part 1--5) in the appendix presents the \gls{Q2Q} matrix for the \gls{EnVis} scenarios.

As a concluding point in this section, it is important to note that these steps are not strictly unidirectional and can provide feedback to earlier stages, as illustrated in Figure \ref{fig:methodology}. The quantification process is also closely tied to the qualitative scenarios, enabling iterative feedback for their refinement. This study applied such feedback to revise the \gls{Q2Q} matrix.

\section{EU EnVis-2060: Initial Quantification} \label{sec:quantitave}

The quantification of the \gls{EnVis} storylines was performed in a two-step approach: first, the qualitative storylines were taken and parametrized using the \gls{Q2Q} matrix, as well as the overall descriptions. Second, the created parameter sets were fed into the open-source energy system model GENeSYS-MOD \cite{loffler_designing_2017}.

\subsection{Parametrization}
For the parametrization, a wide-spread literature research for key data points was conducted. The general trends outlined in the \gls{Q2Q} matrix were then taken to choose the corresponding data points from the literature. An example of this would be, e.g. for capital costs of technologies, where sufficient data exists in literature: following the \gls{Q2Q} matrix, the highest level of innovation and thus cost reductions happen in Go RES, thus the lower end of the capital costs found in the literature were chosen here. This was repeated across the entire scenario spectrum to find appropriate data for each individual data point. Where sufficient data was difficult to find (e.g. for cost projections of fossil fuels towards 2060), other models and/or experts were consulted to give parameter estimations. The full data set, together with the used references can be found on Zenodo \cite{loffler_european_2024}.\\
\begin{figure}
    \centering
    \includegraphics[width=.72\textwidth, trim=20pt 80pt 0pt 40pt]{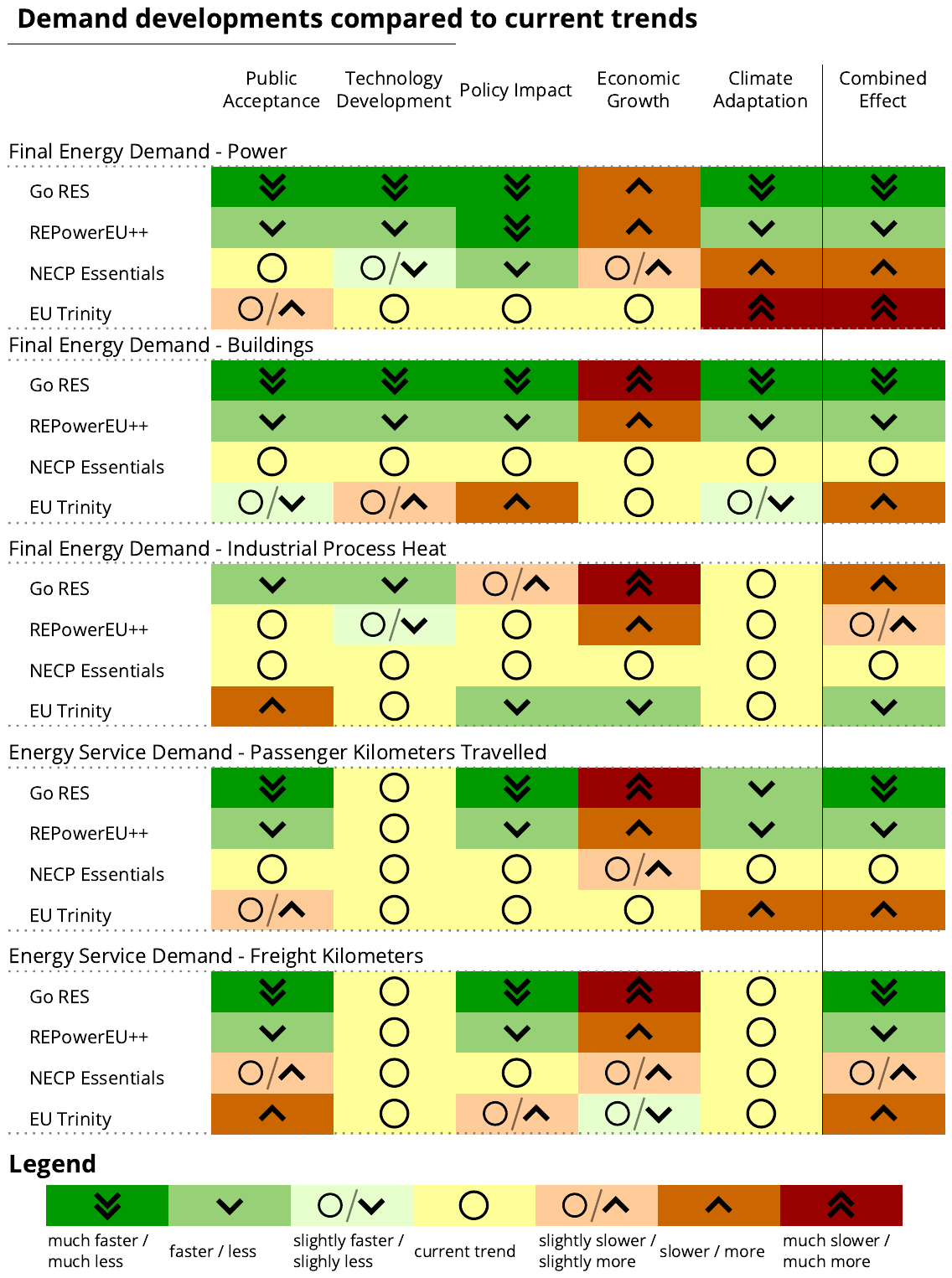}
    \caption{Synthesis of final energy demand developments based on the \gls{Q2Q} matrix.}
    \label{fig:finalenergydemands}
\end{figure}
For purely scenario-specific datapoints, such as demand projections, own assumptions based on the storylines were used. The starting point for each scenario is based on historic data for the year 2018 from statistical sources, while the projections then are constructed using historic trends (giving the baseline development of each final demand sector), combined with the end-use demand specifications from the storylines, as outlined in the \gls{Q2Q} matrix. Figure \ref{fig:finalenergydemands} shows the resulting synthesis for the final demand sectors of GENeSYS-MOD that were then used in the successive model applications. 

\subsection{Energy System Modelling with GENeSYS-MOD}

The parametrization outlined in the previous Subsection was then applied to the Global Energy System Model (GENeSYS-MOD), which is an open-source, multi-energy-carrier, energy system model. It features the sectors electricity, buildings, transport, and industry and is specialized in investigating long-term pathways for the energy system. It does so by optimizing the new-present-value of the energy system towards the future (in our case 2060) for a specified set of regions, while computing the necessary generation capacity additions, energy flows, and flexibility requirements. Since it includes and optimizes all sectors of the energy system simultaneously, while also taking the entire modeled time horizon into account, it is a powerful tool to gain insights into long-term trends of the energy system \cite{loffler_designing_2017}.\\

Some key results of this first quantification of the EU EnVis-2060 scenarios can be found in Table \ref{tab:quantification}. A more detailed scenario quantification spanning all four scenarios will be conducted in the future, but would go beyond the scope of this paper.

\begin{table*}
    \centering
    \caption{Initial quantification of the scenarios \textit{Go RES} and \textit{REPowerEU++} with GENeSYS-MOD.}
    \makebox[\textwidth][c]{%
    \includegraphics[width=1.1\textwidth]{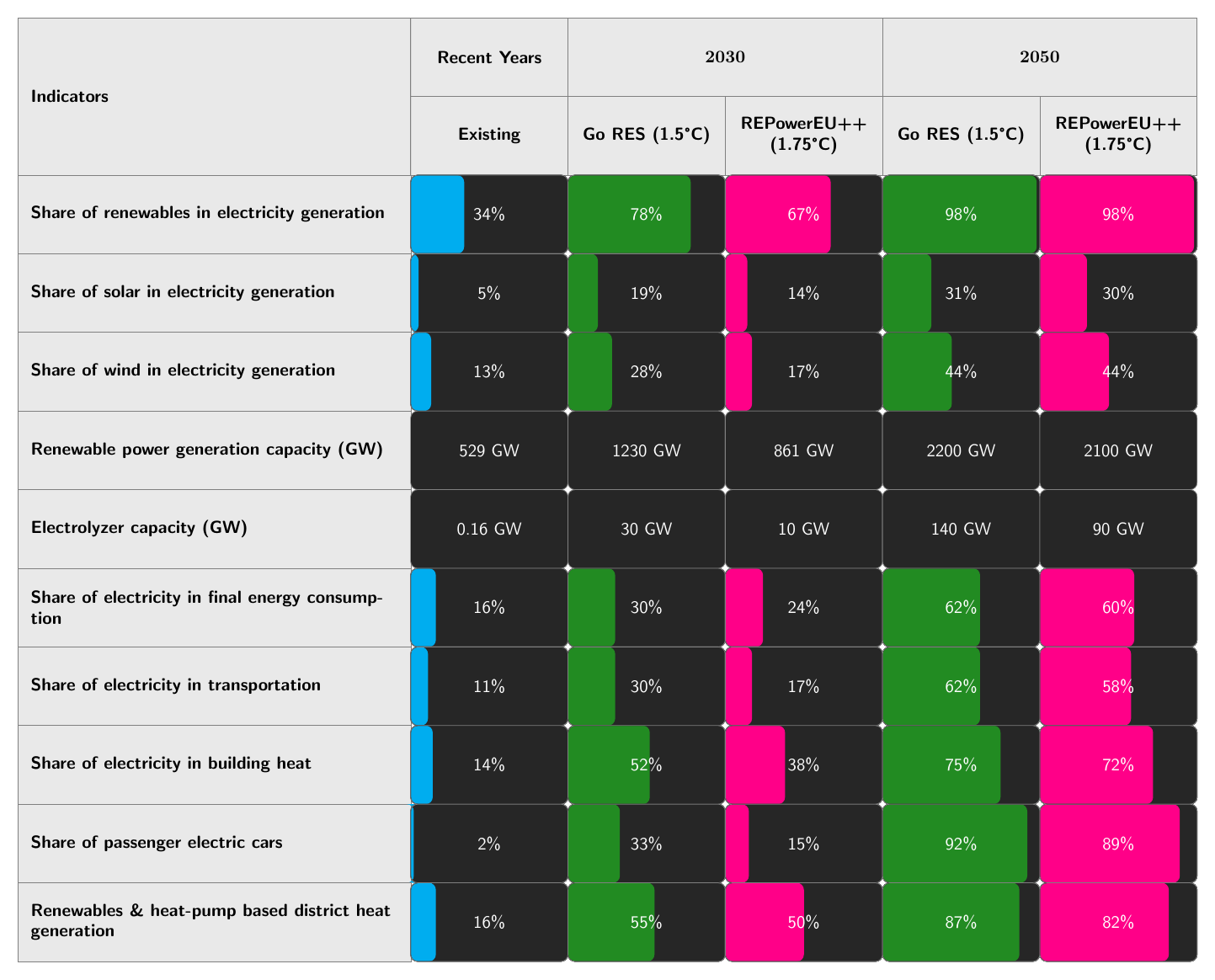}}
    \label{tab:quantification}
\end{table*}

As seen in Table \ref{tab:quantification}, the ambitious climate goals outlined in the \emph{REPowerEU++} and \emph{Go RES} scenarios both require a tremendous effort in terms of the energy transition and integration of renewable energy sources. The major difference between the two scenarios is the speed at which the transition is required in the early years and highlights the large gaps that exist in the current energy infrastructure compared to the needed targets. Especially today's electrolyzer capacities are far below those needed for a successful achieval of the EU-EnVis scenarios.

\section{Concluding Remarks}\label{sec:conclusion}

This paper conducts a comprehensive review of existing energy transition studies and identifies a common conclusion: achieving net-zero GHG emissions by 2050 and the European Green Deal targets requires urgent action and a coordinated, large-scale effort across governments, societies, and industries.
Engaging several experts and relevant stakeholders, the study develops four long-term energy transition scenarios—EU Trinity, NECP Essentials, REPowerEU++, and Go RES—ranging from pessimistic to optimistic. These scenarios outline diverse pathways for Europe’s energy transition through 2060, considering five overarching uncertainties: social, technological, geopolitical, policy, and economic factors. These dimensions serve as broad umbrellas, together shaping specific uncertainties such as the extent of cost reductions in renewable technologies, the policies governing GHG budgets, and the availability of resources and raw materials, all of which are critical for the quantification analysis of energy systems. To manage the state space of the scenarios and facilitate discussions shaping the narratives, the study first focuses on three of these uncertainties, forming a three-dimensional space to locate the scenarios. The other two uncertainties are then incorporated through morphologic analysis.
A key contribution of this work is its emphasis on transparency in scenario development, clarity in assumptions, and transparency in quantification. To this end, the paper introduces the Qualitative to Quantitative (Q2Q) matrix, which systematically details assumptions related to policies, technology and infrastructure, consumption behaviors, independence metrics, and primary energy resources. This structured approach ensures clarity and enhances the connection between qualitative narratives and the quantification process, providing a partially systematized framework for scenario modeling.
Furthermore, all scenarios developed in this study will be openly available and reusable, supporting broader access and fostering future research into Europe’s energy transition.

\section*{Acknowledgement}
The authors would like to thank Sandrine Charousset-Brignol (EDF) for her invaluable support and contributions at all stages of scenario development, as well as Paolo Pisciella (NTNU) for his expert insights. The authors also extend their gratitude to all Man0EUvRE and iDesignRES partners and stakeholders for their collaborative and thoughtful feedback throughout the entire process of this work.

\section*{Conflict of Interest Statement}
The authors declare that the research was conducted in the absence of any commercial or financial relationships that could be construed as a potential conflict of interest.

\section*{Funding}
This research was mainly funded by CETPartnership, the European Partnership under Joint Call 2022 for research proposals, co-funded by the European Commission (GA Nº101069750) and with the funding organisations listed on the CETPartnership website.

Franziska M. Hoffart would like to thank the German Federal Government, the German State Governments, and the Joint Science Conference (GWK) for their funding and support as part of the NFDI4Energy consortium. Funded by the Deutsche Forschungsgemeinschaft (DFG, German Research Foundation) –501865131.

\section*{Declaration of Generative AI and AI-assisted Technologies in the Writing process}

During the preparation of this work the authors used ChatGPT in order to enhance the language and style. After using this tool/service, the authors reviewed and edited the content as needed and take full responsibility for the content of the publication.

\bibliographystyle{elsarticle-num} 
\bibliography{references}

\section*{Appendix A: Brain Storming}
\begin{figure*}[h!]
    \centering
    \makebox[\textwidth][c]{%
    \begin{tikzpicture}
    \path[draw=black, rounded corners] (-220pt, -104pt) rectangle (220pt,104pt);
        \node[] {\includegraphics[width=1.1\textwidth,]{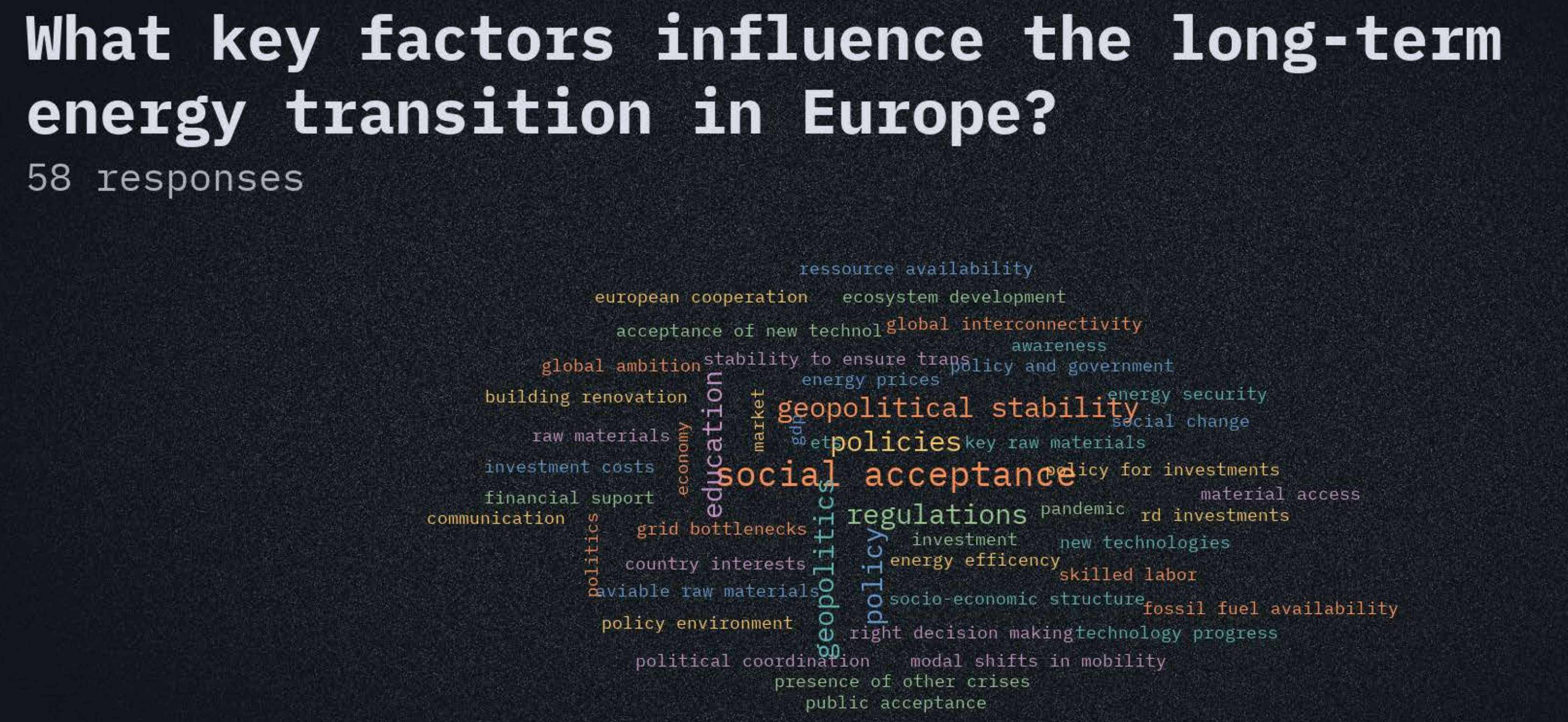}};
    \end{tikzpicture}}
    \caption{Responses to a question during the brainstorming session.}
    \label{fig:brainstorming}
\end{figure*}

Figure \ref{fig:brainstorming} displays the responses collected during a brainstorming session with all partners to a specific question.

\clearpage

\section*{Appendix B: Merging process}
\begin{figure*}[h!]
    \centering
    \makebox[\textwidth][c]{%
    \includegraphics[width=1.1\textwidth]{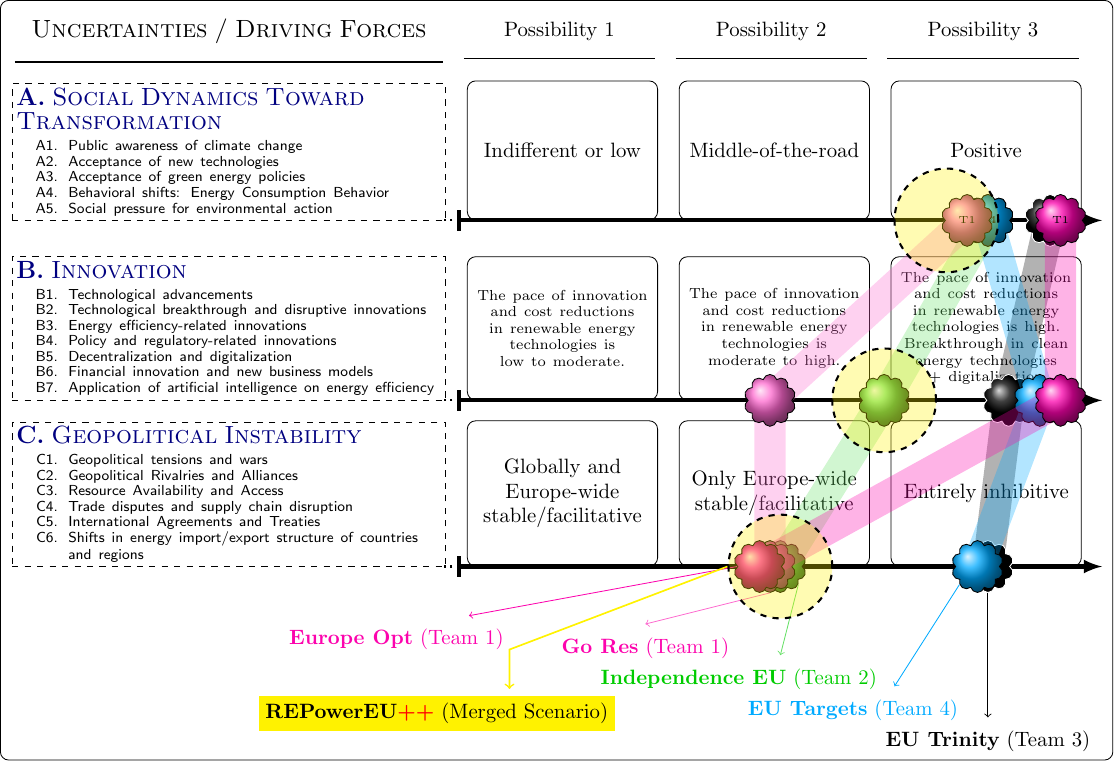}}
    \caption{The merged scenario out of the various proposals.}
    \label{fig:sample_merging}
\end{figure*}

Figure \ref{fig:sample_merging} depicts the REPowerEU++ scenario, which merges elements from a set of closely related scenarios proposed by different groups within the core team.

\section*{Appendix C: Narrative Table}
Table \ref{tab:storyline_table} provides detailed information on the status of various sub-elements in relation to the primary uncertainties for each scenario.

\newgeometry{top=2cm, bottom=2cm, left=1.25cm, right=1.25cm}

\begin{table*}
    \centering
    \caption{ Driving Forces / Uncertainties Table --- Part 1 of 2.}
    \makebox[\textwidth][c]{\includegraphics[page=1]{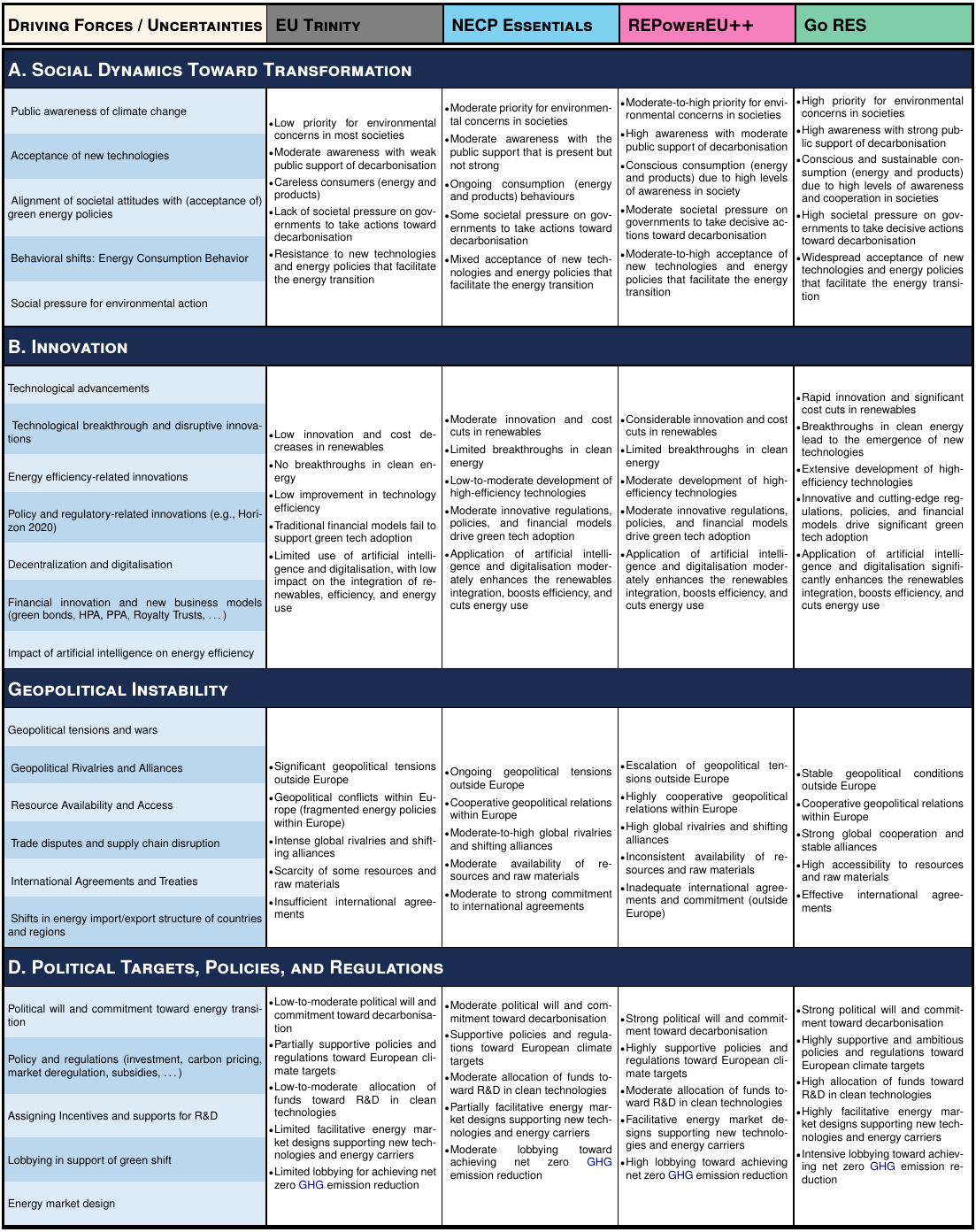}}
    \label{tab:storyline_table}
\end{table*}

\begin{table*}
    \centering
    \caption*{Table \ref*{tab:storyline_table}: Driving Forces / Uncertainties Table --- Part 2 of 2.}
    \makebox[\textwidth][c]{\includegraphics[page=2]{figures/StorylineTable.pdf}}
\end{table*}

\restoregeometry

\section*{Appendix D: \gls{Q2Q} Matrix}

Table \ref{tab:Q2Q} represents the \acrfull{Q2Q} matrix.

\newgeometry{top=2cm, bottom=2cm, left=1.25cm, right=1.25cm}

\begin{table*}
    \centering
    \caption{Q2Q Matrix: Detailing qualitative scenarios to be linked to modelling tools --- Part 1 of 5.}
    \makebox[\textwidth][c]{\includegraphics[page=1]{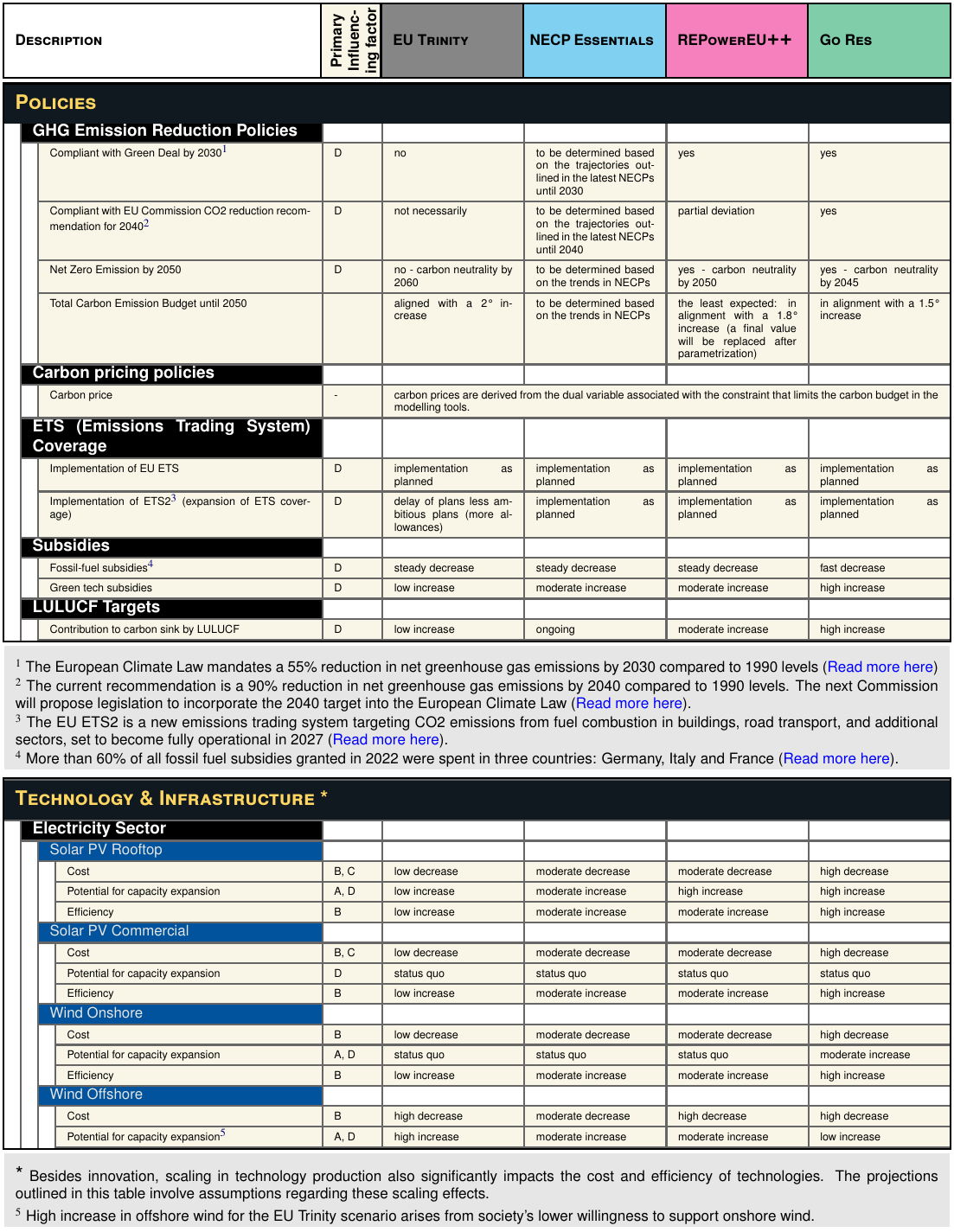}}
    \label{tab:Q2Q}
\end{table*}

\begin{table*}
    \centering
    \caption*{Table \ref*{tab:Q2Q}: Q2Q Matrix: Detailing qualitative scenarios to be linked to modelling tools --- Part 2 of 5.}
    \makebox[\textwidth][c]{\includegraphics[page=2]{figures/Q2Q.pdf}}
\end{table*}

\begin{table*}
    \centering
    \caption*{Table \ref*{tab:Q2Q}: Q2Q Matrix: Detailing qualitative scenarios to be linked to modelling tools --- Part 3 of 5.}    
    \makebox[\textwidth][c]{\includegraphics[page=3]{figures/Q2Q.pdf}}
\end{table*}

\begin{table*}
    \centering
    \caption*{Table \ref*{tab:Q2Q}: Q2Q Matrix: Detailing qualitative scenarios to be linked to modelling tools --- Part 4 of 5.}    
    \makebox[\textwidth][c]{\includegraphics[page=4]{figures/Q2Q.pdf}}
\end{table*}

\begin{table*}
    \centering
    \caption*{Table \ref*{tab:Q2Q}: Q2Q Matrix: Detailing qualitative scenarios to be linked to modelling tools --- Part 5 of 5.}
   \makebox[\textwidth][c]{\includegraphics[page=5]{figures/Q2Q.pdf}}
\end{table*}

\restoregeometry

\end{document}